\documentclass{svjour2}                    
\smartqed  
\usepackage{graphicx}
\usepackage{amsmath}
\usepackage{amssymb}
\usepackage{float}
\usepackage{epsfig}
\usepackage[numbers]{natbib}

\newcommand{\beq}{\begin{equation}}
\newcommand{\eeq}{\end{equation}}
\newcommand{\bea}{\begin{eqnarray}}
\newcommand{\eea} {\end{eqnarray}}

\newcommand{\RR}{{\mathbb R}}
\newcommand{\CC}{{\mathbb C}}
\newcommand{\DD}{{\mathbb D}}

\def\un\a{{\underline\alpha}}

\def\a{{\alpha}}

\newtheorem{defn}{Definition}[section]

\begin{document}

\title{Quantum Mechanics reconstruction from invariance of the laws of nature under tensor composition
}


\author{Florin Moldoveanu
}


\institute{Florin Moldoveanu \at
              Logic and Philosophy of Science Research Group, University of Maryland at College Park \\
              \email{fmoldove@gmail.com}           
}

\date{Received: date / Accepted: date}

\maketitle

\begin{abstract}
Quantum and classical mechanics are derived using four natural physical principles: (1) the laws of nature are invariant under time evolution, (2) the laws of nature are invariant under tensor composition, (3) the laws of nature are relational, and (4) positivity (the ability to define a physical state). Quantum mechanics is singled out by a fifth experimentally justified postulate: nature violates Bell's inequalities.
\end{abstract}

\section{Introduction}
\label{intro}
Quantum mechanics is an extremely successful theory of nature and yet it has resisted all attempts to date to have an intuitive and complete axiomatization. Various attempts were made over time to derive quantum mechanics. For example, Piron was able to show that any propositional lattices respecting orthomodularity, completeness, atomicity and the covering property is isomorphic to the lattice of subspaces of a Hilbert space over some field \cite{PironQM}. 

Recent approaches are using system composition arguments in various forms to recover the quantum mechanics formalism: in an instrumentalist derivation \cite{HardyQM}, in the context of combining experiments sequentially or in parallel \cite{GoyalQM}, or in the context of quantum information \cite{DakicBrukner,MasanesQM}. Other recent attempts are approaching the problem by selecting distinctions between classical and quantum information alone or in combination with composition or other arguments \cite{FuchsQM,ChiribellaQM,ChiribellaDerivation,BarnumWilce,BarnumMullerUdudec}. 

The present approach starts from the observation that quantum and classical mechanics have very similar algebraic mathematical structures centered on observables which play a dual role as observables and generators. In the quantum case one encounters a Jordan-Lie algebra and the corresponding classical mechanics mathematical structure is a Poisson algebra \cite{LandsmanBook}. 

Beside the Jordan-Lie and Poisson algebra formalism, there is another set of axioms introduced by Segal which are obeyed by both quantum and classical mechanics \cite{SegalAxioms}. However, this set of axioms is too general, because Segals' axioms do not demand the algebra to be involutive. It is the involution property of the C*-algebra formulation of quantum mechanics which generates a ``dynamic correspondence'' between observables and generators \cite{AlfsenShultz}.

\subsection{Motivation and approach}
\label{ProjectMotivation}

Quantum mechanics is described by a set of mathematical structures: Hilbert space, the commutator, the symmetrized product of Hermitian operators. When two quantum systems are combined the mathematical formalism remains the same. For example the tensor product of two Hilbert spaces is still a Hilbert space and observables are still described by self-adjoint operators. Two quantum mechanics systems cannot be combined to generate a classical mechanics system. Very few mathematical structures can obey this self-similarity invariance under tensor composition. It turns out that invariance under tensor composition along with additional natural assumptions completely determine all algebraic properties of quantum mechanics. 

To fully recover quantum mechanics we make the transition from mathematics to physics by requiring the ability to generate information and make experimental predictions. Both quantum and classical mechanics are constructively arising out of composition and information considerations and to distinguish them we appeal to experimental evidence.

The initial assumptions are minimal and rooted into experimental concepts: the existence of time and of a configuration space manifold. First the phase space formulation is recovered and then operators on a Hilbert space are built using deformation quantization. In turn this obtains the norm axioms and completely recovers the usual Hilbert space formulation.

\subsection{Outline of the reconstruction project}
\label{ProjectOutline}

There are three parts to the present quantum mechanics reconstruction project:
\begin{enumerate}

\item Extracting essential mathematical structures from classical and quantum mechanics,

\item Deriving those mathematical structures from physical principles,

\item Reconstructing the full formalism of quantum mechanics.

\end{enumerate}

The first part of the approach was completely solved in the 1970s by Emile Grgin and Aage Petersen \cite{GrginCompPaper}. The original motivation was a belief by Bohr (as reported by his personal assistant Aage Petersen) that the correspondence principle has more to reveal. This led to the discovery that for classical and quantum mechanics the dynamics is invariant under tensor composition of two subsystems. On the mathematical side, the Grgin-Petersen approach is identical with the Jordan-Lie algebraic approach to quantum mechanics except for a key difference: it does not include the norm axioms. This may be perceived as a weakness because when one looks at the definition of operator algebras \cite{AlfsenShultz}, one notices separate algebraic and norm properties. However for C*-algebras the norm (which is also used by Jordan-Lie algebras) is unique \cite{ConnesBook} and defined by the spectral radius - an algebraic concept: 

\begin{eqnarray*}\label{SpectralRadius}
{||T||}^2 &=& {\rm Spectral~radius~of~} T^* T \\ \nonumber
&=& \max \left\{|\lambda | ; \lambda \in \CC {\rm ~and~} (T^* T - \lambda 1) {\rm ~is~not~invertible}\right\}.
\end{eqnarray*}

It is therefore conceivable that quantum mechanics can be reconstructed into a fully algebraic framework and we will later see that positivity will be responsible for recovering the norm axioms. 

For the first part we will present a brief overview of the approach and identify the physical interpretation for the essential algebraic structures and relations. 

Next we will derive those essential algebraic identities from three natural physical principles:

\begin{enumerate}

\item The laws of nature are invariant under time evolution, 

\item The laws of nature are invariant under tensor composition, 

\item The laws of nature are relational.
\end{enumerate}

This will lead to three possible solutions: elliptic composition (corresponding to quantum mechanics), parabolic composition (corresponding to classical mechanics), and hyperbolic composition (corresponding to hyperbolic quantum mechanics over split-complex numbers \cite{KhrennikovSegre}). 

The hyperbolic composability solution will be shown to violate a fourth principle: positivity. This means that one cannot construct a state space able to always generate physical non-negative probability predictions. 

The remaining two solutions corresponding to classical and quantum mechanics have two concrete mathematical realizations of the algebraic identities which in the end can be proven to be either: functions on phase space or operators on a Hilbert space. The problem is that we cannot simply assume either a phase or a Hilbert space and we need to derive them. Working in one of the realizations we derive the Hamiltonian formalism for classical and the phase space formalism for quantum mechanics. Then for elliptic composability we will use Berezin deformation quantization \cite{BerezinQuantization} to recover the Hilbert space. From this we can extract the C*-algebra condition:

\begin{equation*}
||x^* x|| = {||x||}^2 ,
\end{equation*}  

\noindent and recover the usual C*-algebra formulation of quantum mechanics. 

In this paper we will not discuss the reconstruction problem in the infinite degree of freedom case (quantum field theory) which is a much harder problem.  

To distinguish between classical and quantum mechanics, there are several options available. Some approaches to deriving quantum mechanics use information theoretical arguments but since quantum and classical mechanics belong to completely disjoint composability classes we can simply appeal to experimental evidence to determine which composability class is selected by nature. We note that it is no longer necessary to find a criteria to distinguish between quantum mechanics and the hypothetical Popescu-Rohrlich (PR) boxes \cite{PR1} because such devices are forbidden by the first four axioms. 

The approach of Grgin and Petersen is categorical in nature and was recently put in the category formalism \cite{AKapustin}. Then quantum mechanics was recovered in the finite dimensional case (spin systems) by an appeal to the Artin-Wedderburn theorem \cite{ArtinWedderburn}.  The advantage of the current approach besides deriving the assumptions of the Grgin-Petersen formalism is that it solves the reconstruction problem for infinite dimensional Hilbert spaces.

In the present approach there is freedom in selecting the number system for quantum mechanics. It can be shown that octonionic quantum mechanics is not a consequence of the current axioms, but all other representations allowed by real Jordan algebras classification corresponding to projective spaces demanded by Piron's result are allowed. For example quaternionic quantum mechanics \cite{AdlerQuaternions} can be understood as complex quantum mechanics subject to a constraint \cite{BengtssonBook}: observables are invariant under time reversal. Whenever multiple representations are possible, all of them give the same predictions. 

The number system representation for quantum mechanics is an open research area which goes beyond the usual real, complex or quaternionic numbers \cite{GrginBook} but we will restrict discussion in this paper mainly to the usual formulation of quantum mechanics using complex numbers. 

Last, we will discuss some implications of the current results and present a list of open problems.

\section{Extracting essential mathematical structures from classical and quantum mechanics}
\label{essential}

Before we give any definitions, let us start with a high level overview. Both quantum and classical mechanics have Hilbert space realizations \cite{ClassicalInHilbertSpace,vonNeuman} and similarly, both have phase space formulations \cite{WignerFunctions}. Grgin and Petersen formalism is called a two-algebra formalism \cite{GrginCompPaper} because it includes two algebraic products, one symmetric and one skew-symmetric. The symmetric product corresponds to observables and is the usual function multiplication in the classical case, and the Jordan product in the quantum case. The skew-symmetric product corresponds to dynamics and is the Poisson bracket for classical mechanics, and the commutator for quantum mechanics. In the Hilbert space formalism the two products act on different spaces linked by a one-to-one map: the space of observables and the space of generators. In the phase space formalism for symplectic manifolds there is a one-to-one map in the cotangent bundle between canonical coordinates. There is also a compatibility condition between the two products which is trivial in the case of classical mechanics, but non-trivial for quantum mechanics. This nontriviality of the compatibility condition is the root cause of quantum superposition and entanglement. 

Let us call the symmetric product $\sigma$ and the skew-symmetric product $\alpha$. For readability, when appropriate, we will use either uppercase ($A, B, C$) or lowercase letters ($f, g, h$) to denote elements from the domain of the products $\sigma$ and $\alpha$, but unless the representation is specified this does not imply that they are operators on a Hilbert space or functions on a phase space and we will treat the products in an abstract way.

For any product $\circ$, the {\em associator}: 
\begin{equation*}
{[A, B, C]}_{\circ} = (A \circ B) \circ C - A \circ (B \circ C)
\end{equation*}
\noindent quantifies the violation of associativity. With those preliminaries the identities respected by quantum and classical mechanics are: Leibniz, Jacobi, Jordan, and a compatibility relation:

\begin{eqnarray*}
Leibniz & A\alpha (B \circ C) = (A\alpha B) \circ C + B \circ (A \alpha C) ,\\
Jacobi & A\alpha (B \alpha C) + C\alpha (A \alpha B) + B\alpha (C \alpha A) = 0 ,\\
Jordan & A \sigma (B \sigma (A \sigma A)) = (A \sigma B) \sigma (A \sigma A) ,\\ 
compatibility & {[ A, B, C] }_{\sigma} + \frac{J^2 \hbar^2}{4} {[A, B, C]}_{\alpha} = 0 ,\\
\end{eqnarray*}

\noindent along with three supplemental properties:

\begin{eqnarray*}
relationality & 1 \alpha A = A \alpha 1 = 0 ,\\
unitality & 1 \sigma A = A \sigma 1 = A ,\\
involution & J \rightarrow (-J) ,\\
\end{eqnarray*}

\noindent where $J^2 = -1$ for quantum mechanics and $J^2 = 0$ for classical mechanics (a third unphysical case corresponding to a hyperbolic quantum mechanics over split-complex numbers \cite{KhrennikovSegre} can be formally obtained by demanding $J^2 = +1$). In all equations above $\circ$ could be either $\alpha$ or $\sigma$.

Please note that in the classical case, from the compatibility condition the product $\sigma$ is associative, which is stronger than just power associativity. Also the Jordan algebras are not required to be formally real ($A^2 + B^2 = 0 \Rightarrow A=B=0$). 

Physically the four identities correspond to time, dynamics, observables and states, and the additional properties are related to ground energy level, and Noether theorem.

\begin{table}[H] 
\caption{From identities to physics}
\label{identitiestophysics}       
\begin{tabular}{ll}
\hline\noalign{\smallskip}
{\em Identity/Property} & {\em Physical interpretation} \\
\noalign{\smallskip}\hline\noalign{\smallskip}
Leibniz & Time\\
Jacobi & Dynamics \\
Jordan & Observables\\
Compatibility & States\\
\noalign{\smallskip}\hline\noalign{\smallskip}
Relationality & Ground energy level does not affect the dynamic\\
\noalign{\smallskip}\hline\noalign{\smallskip}
Unitality & Invariance under tensor composition: Ground \\
          & energy level is invariant under tensor composition\\
\noalign{\smallskip}\hline\noalign{\smallskip}
Involution & Noether theorem: observables are generators of \\
           & kinematic symmetries and an observable which is\\
           & preserved by time evolution generates a dynamical \\
           & continuous symmetry \cite{AKapustin}\\
\noalign{\smallskip}\hline
\end{tabular}
\end{table}

For quantum mechanics, the Hilbert space formulation for the products $\alpha$ and $\sigma$ is:

\begin{eqnarray*}
A\alpha B &=& \frac{J}{\hbar}(AB - BA), \\ \nonumber
A\sigma B &=& \frac{1}{2} (AB + BA) ,
\end{eqnarray*}
\noindent which are the usual commutator and the Jordan product. In (flat) phase space formulation the products $\alpha$ and $\sigma$ are the Moyal and the cosine bracket \cite{MoyalBracket}:

\begin{eqnarray*}
\alpha &=& \frac{2}{\hbar}\sin ( \frac{\hbar}{2}\overleftrightarrow{\nabla}) ,\\ \nonumber
\sigma &=& \cos (\frac{\hbar}{2}\overleftrightarrow{\nabla}) ,
\end{eqnarray*}

\noindent where the operator $\overleftrightarrow{\nabla}$ is defined as follows:

\begin{equation*}
\overleftrightarrow{\nabla} = \sum_{i=1}^{N} [ \overleftarrow{\frac{\partial}{\partial x_i}}\overrightarrow{\frac{\partial}{\partial p_i}} - \overleftarrow{\frac{\partial}{\partial p_i}}\overrightarrow{\frac{\partial}{\partial x_i}}] .
\end{equation*}

It is a straightforward exercise to prove that both realizations respect the four identities and the supplemental properties. Something more can be defined: an associative product beta: $\beta = \sigma \pm \frac{J\hbar}{2} \alpha$. For the Hilbert space representation $\beta = \sigma - \frac{J\hbar}{2} \alpha$ is the usual operator multiplication, while for the phase space formulation $\beta = \sigma + \frac{J\hbar}{2} \alpha$ is the star product:

\begin{equation*}
f \star g = f \sigma g + \frac{J\hbar}{2} f \alpha g = f e^{\frac{J\hbar}{2} \overleftrightarrow{\nabla}} g .
\end{equation*}

The plus and minus choice correspond to either the ordinary multiplication or the reversed multiplication.

Combining all of the above we can have the following definition:

{\defn A composability two-product algebra is a real vector space $\mathfrak{A}_{\RR}$ equipped with two bilinear maps $\sigma$ and $\alpha$ such that the following conditions apply:
\begin{eqnarray*}
\alpha {\rm ~is~a~Lie~algebra} ,\\
\sigma {\rm ~is~a~Jordan~algebra} ,\\
\alpha {\rm ~is~a~derivation~for~}\sigma {\rm ~and~} \alpha ,\\
{[ A, B, C] }_{\sigma} + \frac{J^2 \hbar^2}{4} {[A, B, C]}_{\alpha} = 0 ,
\end{eqnarray*}
where $J \rightarrow (-J)$ is an involution, $1\alpha A = A\alpha 1 = 0$, $1\sigma A = A\sigma 1 = A$, and $J^2 = -1,0,+1$. 
} 

Quantum mechanics corresponds to $J^2 = -1$ (elliptic composability), classical mechanics corresponds to $J^2 = 0$ (parabolic composability), and the unphysical hyperbolic quantum mechanics corresponds to $J^2 = +1$ (hyperbolic composability).

\section{Deriving the composability two-product algebra from physical principles}
\label{composabilityDerivation}

In step one of the quantum mechanics reconstruction program we presented the essential identities respected by quantum and classical mechanics. In step three we will see that the $J^2 = -1$ composability two-product algebra together with positivity is an alternative formulation of complex quantum mechanics because it contains all the information needed to recover the usual Hilbert space formulation. In the current step we will derive the composability two-product algebra from three physical principles: laws of nature are invariant under time evolution, laws of nature are invariant under tensor composition, and laws of nature are relational. 

Invariance of the laws of nature under time evolution is self-evident. Invariance of the laws of nature under tensor composition means that if system A is described by quantum mechanics, and system B is described by quantum mechanics, then the total system $A\otimes B$ is described by quantum mechanics as well. Alternatively, if we can experimentally determine the values of the Planck constant to be $\hbar_A$, $\hbar_B$, and $\hbar_{A\otimes B}$, then all those values are identical \cite{SahooPlanck}. 
 
Laws of nature are relational means that not only kinematics obeys the principle of relativity, but that dynamics is insensitive to constant values as follows: the ground energy level value does not affect the dynamics and the tensor product has a unit in the form of constant functions.

Since the two products $\sigma$ and $\alpha$ are the main mathematical structures in the algebraic formulation of quantum mechanics, it is conceivable that they may depend on the physical system. We start with two physical systems $A$ and $B$. Suppose the products $\alpha_A, \sigma_A$ (one skew-symmetric and one symmetric) apply to system $A$, and correspondingly $\alpha_B, \sigma_B$ apply to system $B$. By tensor composition and symmetry property, the total system $T = A\otimes B$ is described by the following products:

\begin{eqnarray*}
(f_A \otimes f_B) \alpha_{A\otimes B} (g_A \otimes g_B) &=& a {(f \alpha g)}_A\otimes {(f \sigma g)}_B + b {(f \sigma g)}_A\otimes {(f \alpha g)}_B ,\\
(f_A \otimes f_B) \sigma_{A\otimes B} (g_A \otimes g_B) &=& c {(f \sigma g)}_A\otimes {(f \sigma g)}_B + d {(f \alpha g)}_A\otimes {(f \alpha g)}_B .
\end{eqnarray*}  

Can we then determine the four values of the parameters $a, b, c, d$? What Grgin and Petersen originally found \cite{GrginCompPaper} is that assuming $\alpha$ to be a Lie algebra and a derivation to $\sigma$ along with the invariance of the laws of nature under tensor composition demands:

\begin{eqnarray*}
a = b = c = 1 ,\\
d = \frac{J^2 \hbar^2}{4} .
\end{eqnarray*}

\noindent Also it follows that:
\begin{eqnarray*}
\sigma {\rm ~is~a~Jordan~algebra} ,\\
{[ f, g, h] }_{\sigma} + \frac{J^2 \hbar^2}{4} {[f, g, h]}_{\alpha} = 0 .
\end{eqnarray*}

\noindent The natural questions to ask is to what extent we can generalize this approach, and what are the minimal requirements needed to derive the four identities: Leibniz, Jacobi, Jordan, and compatibility together with the symmetry properties for $\alpha$ and $\sigma$?

It is the goal of step two of the quantum mechanics reconstruction program to derive the products $\alpha$ and $\sigma$, their symmetry properties, the involution, Leibniz, relationality, unitality, Jacobi, Jordan, and compatibility condition using only: invariance of the relational laws of nature under time evolution and invariance of the relational laws of nature under tensor composition.

\subsection{Invariance of the relational laws of nature under time evolution}
\label{InvarianceTime}

The algebraic approach to quantum mechanics was originally introduced due to the mathematical difficulties of quantum field theory, in particular the lack of a Hilbert space for certain problems. Citing Emch: ``The basic principle of the algebraic approach is to avoid starting with a specific Hilbert space scheme and rather to emphasize that the {\em primary objects} of the theory are the fields (or the observables) considered as purely algebraic quantities, together with their linear combinations, products, and limits in the appropriate topology.'' \cite{EmchBook}. 

While we work in the algebraic paradigm, we start deriving the composability two-product algebra even more general without assuming the existence of any algebraic products.

From an experimental point of view, we require the existence of time and a configuration space manifold $Q$. At a point $p\in Q$ one can define a tangent space $T_p Q$ and a cotangent space ${T_p}^{*} Q$. From this we form the cotangent bundle manifold $M$. For time evolution we will assume that there are some $C^{\infty}$ functions over $M$ for which there is a way to generate a vector field out of them ($C^{\infty} (M) \rightarrow {\rm Vect} (M)$), and from now on we will restrict the domain of discussion only to those functions. Although other kinds of time evolution are possible (including stochastic time evolution), we are not considering them here.

Let time evolution be represented by a one parameter group of transformations $\phi$ defined as follows:

\begin{equation*}
\phi : M \times \mathbb{R} \rightarrow M ,
\end{equation*}

\noindent with $\phi(x, 0) = x$ and $\phi(\phi(x,t),s)=\phi(x, t+s)$.

Suppose that there is an unspecified family of local operations $\{ \circ \}$ which describe the laws of nature (for example Poisson bracket, Jordan product, commutator, etc). Introducing the notation: $\phi_t (x) \equiv \phi(x,t)$, the invariance of the laws of nature under time evolution reads:

\begin{equation*}
(g \circ h) (\phi_{\Delta t}(p)) =  g (\phi_{\Delta t}(p)) \circ h (\phi_{\Delta t}(p)) ,
\end{equation*}

\noindent with $p \in M$ a point in the manifold $M$ and $g,h$, functions defined in the neighborhood of $p$. In other words, we demand the existence of a universal local morphism which preserves all algebraic relationships under time translation.

If $X^i_f$ is a vector field arising out of a function $f$ (corresponding to a particular time evolution), define $\mathcal{T}_{f_\epsilon}$:

\begin{equation*}
\mathcal{T}_{f_\epsilon} = I + \epsilon X_f^i \frac{\partial}{\partial u^i} ,
\end{equation*}

\noindent with $I$ the identity operator and  $u^i$ the coordinate set in a local $\mathbb{R}^{2n}$ chart covering the point $p\in M$. 

Because $f$ is in one-to-one correspondence with $X_f^i$ we can introduce a time translation transformation $T_{f_\epsilon}$ and a product $\alpha \in \{ \circ \} $ between a distinguished $f$ and any $g$ as follows:

\begin{equation*}
f \alpha g = \lim_{\epsilon \rightarrow 0} \frac{g - T_{f_\epsilon} g}{\epsilon} ,
\end{equation*}

\noindent which is the Lie derivative of $g$ along the vector field generated by $f$ corresponding to a particular time evolution. Equivalently $T_{f_\epsilon} = (I - \epsilon f \alpha \cdot)$.

We generalize the product $\alpha$ for all $f$'s and $g$'s by repeating the argument for all conceivable dynamics. To make sure the domains of $f$ and $g$ are identical and well behaved, in case of pathologies, we can restrict the set of $g$ to the span of all possible $f$.

Then the invariance of the laws of nature under time evolution can be expressed as:

\begin{equation*}
T_{f_\epsilon} (g \circ h) = [T_{f_\epsilon} g] \circ [T_{f_\epsilon} h] ,
\end{equation*}

\noindent which to first order in $\epsilon$ implies a left Leibniz identity:

\begin{equation*}
f \alpha (g \circ h) = (f \alpha g) \circ h + g \circ (f \alpha h) .
\end{equation*}

From the relational property of the laws of nature, we demand for any $f$ that  $1\alpha f = 0$. Also we have $f\alpha 1 = 0$ from the definition of the Lie derivative. $\alpha$ will be shown later to be the usual commutator in quantum mechanics (or the Poisson bracket in classical mechanics) but it has no skew-symmetry property yet.

\subsection{Invariance of the relational laws of nature under tensor composition}
\label{InvarianceTensor}

In this section we will follow the Grgin-Petersen approach \cite{GrginCompPaper} in considering a ``composition class'' $\mathcal{U}=\mathcal{U}(M, \otimes, \mathbb{R}, \alpha, \cdots )$. The class will be later updated to a commutative monoid after commutativity and associativity properties will be proved. The composition class $\mathcal{U}$ has a unit element, the real numbers field $\mathbb{R}$ understood as the set of constant functions. This is a consequence of the relational nature of the laws of nature because constant functions on $M$ have no physical consequences. Formally: $\mathcal{U}\otimes\mathbb{R} = \mathcal{U} = \mathbb{R} \otimes \mathcal{U}$. 

First it can be shown that the product $\alpha$ is not enough and this demands the existence of a second product $\sigma \in \{ \circ \}$.

\subsubsection{The existence of a second product}
\label{SecondProduct}

Here we show that it is impossible to have only one nontrivial product in the composition class. We start from the existence of a unit element for the composition class and we pick $1 \in \mathbb{R}$ understood as a constant function. The existence of a composition class unit demands:

\begin{equation*}
(f \otimes 1) \alpha_{12} (g \otimes 1) = (f \alpha g)\otimes 1 = (f \alpha g) , 
\end{equation*}

\noindent with $\alpha_{12}$ the product $\alpha$ in a bipartite case. Invariance of the laws of nature under composability demands the bipartite products to be built out of the products listed in the composition class. 

Supposing that only a product $\alpha$ exists, $\alpha_{12}$ must be of the form: 

\begin{equation*}
(f \otimes 1) \alpha_{12} (g \otimes 1) = a (f \alpha g) \otimes (1 \alpha 1) ,
\end{equation*}
\noindent but this is zero because $(1 \alpha 1) = 0$.

As such we can have only trivial products $\alpha$. Only by adding another product $\sigma$ we can construct non-trivial mathematical structures.

\subsubsection{The fundamental bipartite relationship}
\label{bipartiteRel}

Let us first observe that two products $\alpha$ and $\sigma$ can always be renormalized by change of units. The overall term of $\frac{\hbar^2}{4}$ is introduced only to agree with the usual product realizations. It is more convenient to work in a convention where $\hbar = 2$ and only the $J^2 = -1, 0, +1$ term remains. 

Originally the fundamental relationship was obtained \cite{GrginCompPaper} with the additional assumptions of symmetry properties but this is not necessary. At this time we do not assume any symmetry or skew-symmetry properties.

We start by considering four functions $f_1 , f_2 , g_1 , g_2$  over the manifold $M$ at a point $p$. By invariance under composability, the most general way to construct the products $\alpha$ and $\sigma$ in a bipartite system is as follows:

\begin{eqnarray*}
(f_1 \otimes f_2) \alpha_{12} (g_1 \otimes g_2) = a (f_1 \alpha g_1) \otimes (f_2 \alpha g_2) + \\ \nonumber
b (f_1 \alpha g_1) \otimes (f_2 \sigma g_2) + c (f_1 \sigma g_1) \otimes (f_2 \alpha g_2) + \\ \nonumber
d (f_1 \sigma g_1) \otimes (f_2 \sigma g_2) , 
\end{eqnarray*}

\begin{eqnarray*}
(f_1 \otimes f_2) \sigma_{12} (g_1 \otimes g_2) = x (f_1 \alpha g_1) \otimes (f_2 \alpha g_2) + \\ \nonumber
y (f_1 \alpha g_1) \otimes (f_2 \sigma g_2) + z (f_1 \sigma g_1) \otimes (f_2 \alpha g_2) + \\ \nonumber
w (f_1 \sigma g_1) \otimes (f_2 \sigma g_2) . 
\end{eqnarray*}

The strategy is to use the existence of the unit of the composition class to determine the coefficients $a,b,c,d,x,y,z,w$. For convenience we also want to normalize the definition of product $\sigma$ such that $1 \sigma f = f \sigma 1 = f$. This can always be done because the constant function $1$ does not affect the dynamic and therefore $1\sigma f = m f$, $f \sigma 1 = n f$. The parameters $m,n \ne 0$ because otherwise we have a trivial product $\alpha$. 

Now we can use the freedom to choose appropriate functions related to the composition class unit. We start by picking the constant functions $f_1 = g_1 = 1 \in \mathbb{R}$ while using $\mathbb{R} \otimes \mathcal{U} = \mathcal{U}$ and $1\alpha 1 = 0$. Under this substitution, in $\alpha_{12}$ only terms corresponding to the $c$ and $d$ coefficients survive and this demands $c=1$ and $d = 0$. Similarly, for $\sigma_{12}$ this demands $z=0$ and $w=1$.

Doing the same thing by picking $f_2 = g_2 = 1 \in \mathbb{R}$ results in $b=1$ and $y=0$. In shorthand notation:

\begin{equation*}
\alpha_{12} = \alpha \sigma + \sigma \alpha + a \alpha \alpha ,
\end{equation*}
and
\begin{equation*}
\sigma_{12} = \sigma \sigma + x \alpha \alpha .
\end{equation*}

Now we will prove that $a=0$. To do this we will use the Leibniz identity on a bipartite system:

\begin{eqnarray*}
(f_1 \otimes f_2) \alpha_{12} [(g_1 \otimes g_2) \alpha_{12} (h_1 \otimes h_2)]=\\ \nonumber
[(f_1 \otimes f_2) \alpha_{12} (g_1 \otimes g_2)] \alpha_{12}(h_1 \otimes h_2) +\\ \nonumber
(g_1 \otimes g_2) \alpha_{12} [(f_1 \otimes f_2)\alpha_{12} (h_1 \otimes h_2)] .
\end{eqnarray*}

Substituting the expression for $\alpha_{12}$ and tracking only the ``$a$'' terms meaning ignoring any terms involving the $\sigma$ product (because $\alpha$ is a linear product) we obtain:

\begin{eqnarray*}
a^2 [f_1 \alpha (g_1 \alpha h_1)] \otimes [f_2 \alpha (g_2 \alpha h_2)] = \\ \nonumber
a^2 [(f_1 \alpha g_1 )\alpha h_1)] \otimes [(f_2 \alpha g_2 ) \alpha h_2)] + \\ \nonumber
a^2 [g_1\alpha (f_1 \alpha h_1)] \otimes [g_2 \alpha (f_2\alpha h_2)] .
\end{eqnarray*}

Applying the Leibniz identity again on the right hand side and canceling terms yields:

\begin{eqnarray*}
a^2 \{ [(f_1 \alpha g_1) \alpha h_1]\otimes [g_2 \alpha (f_2 \alpha h_2)] + \\ \nonumber
[g_1 \alpha (f_1 \alpha h_1)] \otimes [(f_2 \alpha g_2) \alpha h_2] \} = 0 ,
\end{eqnarray*}

\noindent which is valid for all $f_1 , f_2 , g_1 , g_2$ and hence $a=0$.

In the end we have the following fundamental relations:
\begin{eqnarray*}
\alpha_{12} &=& \alpha_1 \sigma_2 + \sigma_1 \alpha_2 ,\\
\sigma_{12} &=& \sigma_1 \sigma_2 + x \alpha_1 \alpha_2 , 
\end{eqnarray*}

\noindent where $x$ can be normalized to be either $-1, 0, +1$.

Please note the formal similarity with complex number multiplication when $x=-1$ which corresponds to quantum mechanics. 

\subsubsection{The skew-symmetry of the product $\alpha$}
\label{skewsymmetry}

Proving that the product $\alpha$ is skew-symmetric: $f\alpha g = -g\alpha f$ is essential for recovering the Hamiltonian formalism. The basic strategy is to use the Leibniz identity for a bipartite system. Writing down the bipartite Leibniz identity:

\begin{equation*}
f_{12} \alpha_{12} (g_{12} \alpha_{12} h_{12}) = g_{12} \alpha_{12} (f_{12} \alpha_{12} h_{12}) + (f_{12} \alpha_{12} g_{12}) \alpha_{12} h_{12} ,
\end{equation*}

\noindent we observe that on the two right hand side terms $f$'s and $g$'s appear in reverse order and we will want to take advantage of this by carefully choosing the bipartite functions. We select $g_1 = 1 = h_2$ and expand the equation above using the fundamental bipartite relation for product $\alpha$.

Expanding the left hand side we get:

\begin{eqnarray*}
(f_1 f_2)\alpha_{12}[{(g \alpha h)}_1{(g \sigma h)}_2 + {(g \sigma h)}_1{(g \alpha h)}_2] =\\ \nonumber
{(f \alpha (g \alpha h))}_1 {(f \sigma (g \sigma h))}_2 +
{(f \sigma (g \alpha h))}_1 {(f \alpha (g \sigma h))}_2 + \\ 
{(f \alpha (g \sigma h))}_1 {(f \sigma (g \alpha h))}_2 +
{(f \sigma (g \sigma h))}_1 {(f \alpha (g \alpha h))}_2 , 
\end{eqnarray*}

\noindent but this is identically zero because in the first two terms  $g_1 = 1$ and in the last two terms $h_2 = 1$.

The first term on the right hand side expands to:

\begin{eqnarray*}
(g_1 g_2)\alpha_{12}[{(f \alpha h)}_1{(f \sigma h)}_2 + {(f \sigma h)}_1{(f \alpha h)}_2] =\\ \nonumber
{(g \alpha (f \alpha h))}_1 {(g \sigma (f \sigma h))}_2 +
{(g \sigma (f \alpha h))}_1 {(g \alpha (f \sigma h))}_2 + \\
{(g \alpha (f \sigma h))}_1 {(g \sigma (f \alpha h))}_2 +
{(g \sigma (f \sigma h))}_1 {(g \alpha (f \alpha h))}_2 . 
\nonumber
\end{eqnarray*}

In this expression the first and third term vanishes because $g_1 = 1$, and the last term vanish because $h_2 =1$. Because $g_1$ and $h_2$ are units for the product $\sigma$, the overall remaining term is:

\begin{equation*}
{(f \alpha h)}_1{(g \alpha f)}_2 .
\end{equation*}

Finally, the first term on the right hand side expands to:

\begin{eqnarray*}
[{(f \alpha g)}_1{(f \sigma g)}_2 + {(f \sigma g)}_1{(f \alpha g)}_2]\alpha_{12}(h_1 h_2) =\\ \nonumber
{((f \alpha g) \alpha h)}_1 {((f \sigma g) \sigma h)}_2 + 
{((f \alpha g) \sigma h)}_1 {((f \sigma g) \alpha h)}_2 + \\
{((f \sigma g) \alpha h)}_1 {((f \alpha g) \sigma h)}_2 +
{((f \sigma g) \sigma h)}_1 {((f \alpha g) \alpha h)}_2 .  
\nonumber
\end{eqnarray*}

In this expression the first two terms vanish because $g_1 = 1$, and the last term vanishes because $h_2 =1$. Because $g_1$ and $h_2$ are units for the product $\sigma$, the overall remaining term is:

\begin{equation*}
{(f \alpha h)}_1{(f \alpha g)}_2 .
\end{equation*}

Putting it all together yields:

\begin{equation*}
0 = {(f \alpha h)}_1[{(f \alpha g)}_2 + {(g \alpha f)}_2] ,
\end{equation*}
 
\noindent which is valid for any arbitrary ${(f \alpha h)}_1$ terms. Hence:

\begin{equation*}
f \alpha g = -g \alpha f ,
\end{equation*}

\noindent and the skew-symmetry of the product $\alpha$ is proved.

\subsubsection{The symmetry of the product $\sigma$}
\label{symmetryProp}

To prove that $f\sigma g = g \sigma f$ one can use a similar approach with the one above by picking $h_1 = 1$ instead. However there is a shorter proof by using the fundamental relationship for $\alpha_{12}$ and the just proved skew-symmetry of $\alpha$.

We start from the bipartite expression for the product $\alpha$:

\begin{equation*}
(f_1 f_2) \alpha_{12} (g_1 g_2) = {(f\alpha g)}_1 {(f\sigma g)}_2 + {(f\sigma g)}_1 {(f\alpha g)}_2 .
\end{equation*}

This is also equal with:

\begin{equation*}
-(g_1 g_2) \alpha_{12} (f_1 f_2) = -{(g\alpha f)}_1 {(g\sigma f)}_2 - {(g\sigma f)}_1 {(g\alpha f)}_2 ,
\end{equation*} 

\noindent and

\begin{equation*}
-(g_1 g_2) \alpha_{12} (f_1 f_2) = {(f\alpha g)}_1 {(g\sigma f)}_2 + {(g\sigma f)}_1 {(f\alpha g)}_2 .
\end{equation*}

We therefore have:

\begin{equation*}
{(f\alpha g)}_1 {[(f\sigma g) - (g\sigma f)]}_2 + {[(f\sigma g) - (g\sigma f)]}_1{(f\alpha g)}_2 = 0 .
\end{equation*}

Suppose now that we pick the functions $f$ and $g$ such that ${(f\alpha g)}_1 \neq 0$ and ${(f\alpha g)}_2 \neq 0$. We then have:

\begin{equation*}
1 \otimes \frac{{[(f\sigma g) - (g\sigma f)]}_2}{{(f\alpha g)}_2} + \frac{{[(f\sigma g) - (g\sigma f)]}_1}{{(f\alpha g)}_1} \otimes 1 = 0 .
\end{equation*}

The only way system $1$ value can be equal with system $2$ value is if both expressions are equal with a constant $c$:

\begin{equation*}
1 \otimes c + c \otimes 1 = 0 .
\end{equation*}

However by using the identity property for the tensor product this means that $c+c=0$ and hence $c=0$. In turn this demands the symmetry of the product $\sigma$: $(f\sigma g) = (g\sigma f)$.

\subsubsection{The Jacobi, Jordan, and the compatibility relations}
\label{JacobiJordan}

The product $\alpha$ is linear in the second term because $(f \alpha \cdot)$ is a derivation, is skew-symmetric, and respects the Leibniz identity:

\begin{equation*}
f \alpha (g \alpha h) = (f \alpha g) \alpha h + g \alpha (f \alpha h) .
\end{equation*}

By the skew-symmetry property we get:

\begin{equation*}
f \alpha (g \alpha h) = -h \alpha (f \alpha g) - g \alpha (h \alpha f) ,
\end{equation*}

\noindent which is the Jacobi identity. Hence $\alpha$ is a Lie algebra. 

The proof of the compatibility relation was first obtained by Grgin and Petersen \cite{GrginCompPaper} (using the assumptions of the symmetry of the product $\sigma$, the skew-symmetry of the product $\alpha$, the Jacobi identities, and the fundamental bipartite relations). Because the proof is rather long and not new, we will only sketch it here for completeness sake. Grgin and Petersen start from the bipartite Jacobi identity:

\begin{equation*}
\sum_{\rm cycl} (f_1 f_2)\alpha_{12} ((g_1 g_2)\alpha_{12}(h_1 h_2)) = 0 .
\end{equation*}

After expansion and usage of the Leibniz identity, it becomes:

\begin{equation*}
\sum_{\rm cycl} {(f \sigma (g \sigma h))}_1 {(f \alpha (g \alpha h))}_2 + {(f \alpha (g \alpha h))}_1 {(f \sigma (g \sigma h))}_2 = 0 .
\end{equation*}

Adding it to a copy of itself but with $g_1$ and $h_1$ interchanged results in:

\begin{equation*}
\{[f, g, h]_{\sigma} + [f, h, g]_{\sigma}\}_1 \{f \alpha (g \alpha h)\}_2 = 
\{(g \alpha (h \alpha f)) + (h \alpha (g \alpha f))\}_1 \{[h, f, g]_{\sigma}\}_2 .
\end{equation*}

This implies a relation of proportionality:

\begin{equation*}
(f \alpha (g \alpha h)) = k [h, f, g]_\sigma .
\end{equation*}

Using the Jacobi identity on the left hand side, it yields the compatibility relationship where $k$ can be normalized to: $+1, 0, -1$. The remaining part of the proof is establishing the relation between $k$ and $J^2$ which occurs in the bipartite expansion of the product $\sigma_{12}$. To this aim Grgin and Petersen use the bipartite Leibniz identity to expand:

\begin{equation*}
(f_1 f_2)\alpha_{12} ((g_1 g_2)\sigma_{12}(h_1 h_2)) ,
\end{equation*}
\noindent and working along similar lines as above they derive a proportionality property which this time involves $J^2$. In the end the compatibility identity is obtained:

\begin{eqnarray*}
{[ f, g, h] }_{\sigma} + J^2 {[f, g, h]}_{\alpha} = 0 .
\end{eqnarray*}

The Jordan identity is a straightforward consequence of the compatibility identity when $f,g,h$ are chosen to be $A, B, A\sigma A$ respectively. With this choice the $\alpha$ associator is zero:

\begin{eqnarray*}
[A, B, A^2]_\alpha = (A \alpha B) \alpha (A^2) - A \alpha (B \alpha A^2) = \\
(A \alpha B) \alpha (A^2) - (A \alpha B) \alpha (A^2) - B \alpha (A \alpha A^2) = 0 .
\end{eqnarray*}

The last term $B \alpha (A \alpha A^2)$ is zero because $A \alpha A^2 = A \alpha (A \sigma A) = (A\alpha A) \sigma A + A \sigma (A \alpha A) = 0$. Hence from the compatibility relationship it yields: $[A, B, A^2]_\sigma = 0$ which is another formulation of the Jordan identity (power associativity).

\subsubsection{The associative product} 
\label{associativeproduct}

At this point all the properties describing the composability two-product algebras are obtained from the invariance of the laws of nature under time evolution along with the invariance of the laws of nature under tensor composition and the relational nature of the dynamic. To arrive at states and transition probabilities, one needs an additional ingredient, an associative multiplication $\beta = \sigma \pm \frac{J\hbar}{2} \alpha$. 

Associativity follows from the associator property of the composability two-product algebra. However each product appears twice and the proof is not obvious.

Let us compute the associator ${[A, B, C]}_{\beta} = (A \beta B) \beta C - A \beta ( B \beta C)$ using the definition of $\beta$:

\begin{eqnarray*}
{[A, B, C]}_{\beta} = (A \sigma B \pm \frac{J \hbar}{2} A \alpha B) \beta C  - A \beta (B \sigma C \pm \frac{J \hbar}{2} B \alpha C)\\ \nonumber
=  (A \sigma B) \sigma C \pm \frac{J \hbar}{2} (A \sigma B ) \alpha C \pm \frac{J \hbar}{2} (A \alpha B) \sigma C + \frac{J^2 \hbar^2}{4} (A \alpha B) \alpha C \\ \nonumber
- A \sigma (B \sigma C) \mp \frac{J \hbar}{2} A \sigma (B \alpha C) \mp \frac{J \hbar}{2} A \alpha (B \sigma C) - \frac{J^2 \hbar^2}{4} A \alpha (B \alpha C) \\ \nonumber
= {[A, B, C]}_{\sigma} + \frac{J^2 \hbar^2}{4} {[A, B, C]}_{\alpha} \\ \nonumber
\pm \frac{J \hbar}{2} \{(A \sigma B) \alpha C + (A\alpha B) \sigma C - A \sigma (B \alpha C) - A\alpha (B \sigma C)\} = 0 .
\end{eqnarray*}

In the last line the terms cancel after using the Leibniz rule for $A\alpha (B\sigma C)$ and $(A \sigma B) \alpha C$.

Because $\beta$ is an associative product and $\sigma$ corresponds to its real part, the Jordan algebra of observables $\sigma$ cannot be special. Hence no octonionic quantum mechanics is possible in the current approach. Later on positivity will restrict the Jordan algebras to real Jordan algebras and this in turn will constrain the allowed number systems for quantum mechanics. We will briefly show an example of a quantum mechanics formulation over a number system different than reals, complex numbers, or quaternions. This number system corresponds to a spin factor and leads to Dirac equation and spinors.

\subsubsection{The involution}
\label{Involution}

For the elliptic and hyperbolic composability cases from the fundamental bipartite relation we see that the domain of the products $\alpha$ and $\sigma$ must be identical. This gives rise to a one-to-one involution map between observables and generators known as ``dynamic correspondence''. In the parabolic case because $J^2 = 0$ the involution is no longer a mathematical necessity and one can encounter odd-dimensional Poisson manifolds. 
  
\subsubsection{The commutative monoid}
\label{commutativemonoid}

Last in this section, we will derive the properties of the composition class $\mathcal{U}(M, \otimes, \mathbb{R}, \alpha, \sigma, J)$ and show that it is associative and commutative. Together with its unit, the composition class becomes a commutative monoid (monoidal category in category language). This may look like an unimportant mathematical fact, but it has deep implications for the collapse postulate as we will show later.

The tensor product $\otimes$ is already commutative and associative, and all that remains to be proven are the following identities:

\begin{eqnarray*}
\sigma_{12} &=& \sigma_{21} ,\\
\alpha_{12} &=& \alpha_{21} ,\\
\sigma_{(12)3} &=& \sigma_{1(23)} ,\\
\alpha_{(12)3} &=& \alpha_{1(23)} .\\
\end{eqnarray*}

The first two properties follow from the commutativity of $\otimes$:

\begin{eqnarray*}
\sigma_{12} = \sigma_1 \otimes \sigma_2 + J \alpha_1 \otimes \alpha_2 =
\sigma_2 \otimes \sigma_1 + J \alpha_2 \otimes \alpha_1 = \sigma_{21} ,\\
\alpha_{12} = \alpha_1 \otimes \sigma_2 + \sigma_1 \otimes \alpha_2 =
\sigma_2 \otimes \alpha_1 + \alpha_2 \otimes \sigma_1 = \alpha_{21} .\\
\end{eqnarray*}

The last two identities are straightforward double application of the fundamental bipartite relationships:

\begin{eqnarray*}
\sigma_{1(23)} =& \sigma_1 \sigma_{23} + J \alpha_1 \alpha_{23} &= \sigma_1 \sigma_2 \sigma_3 + J \sigma_1 \alpha_2 \alpha_3 + J \alpha_1 \alpha_2 \sigma_3 + J \alpha_1 \sigma_2 \alpha_3 ,\\
\sigma_{(12)3} =& \sigma_{12} \sigma_3 + J \alpha_{12} \alpha_3 &= \sigma_1 \sigma_2 \sigma_3 +  J \alpha_1 \alpha_2 \sigma_3 + J \alpha_1 \sigma_2 \alpha_3 + J \sigma_1 \alpha_2 \alpha_3 ,\\
\alpha_{1(23)} =& \alpha_1 \sigma_{23} + \sigma_1 \alpha_{23} &= \alpha_1 \sigma_2 \sigma_3 + J \alpha_1 \alpha_2 \alpha_3 + \sigma_1 \alpha_2 \sigma_3 + \sigma_1 \sigma_2 \alpha_3 ,\\
\alpha_{(12)3} =& \alpha_{12} \sigma_3 + \sigma_{12} \alpha_3 &= \alpha_1 \sigma_2 \sigma_3 + \sigma_1 \alpha_2 \sigma_3 + \sigma_1 \sigma_2 \alpha_3 + J \alpha_1 \alpha_2 \alpha_3 .\\
\end{eqnarray*}

\subsection{The relational property of the dynamic}
\label{relational}

To summarize, the relational property of the dynamic was used in parallel with the invariance of the laws of nature under time evolution and the invariance of the laws of nature under tensor composition. Once we introduced the products $\alpha$ and $\otimes$, we used the relational property of the dynamic to derive two properties: $1\alpha f = 0$ and the existence of the unit for the tensor product: $\mathcal{U}\otimes\RR = \mathcal{U} = \RR \otimes \mathcal{U}$. Both of those properties were essential in deriving the composability two-product algebra. 

\section{Reconstructing the full formalism of quantum mechanics}
\label{fullReconstruction}

At this point in the reconstruction program we have derived the composability two-product algebra from physical principles and we are ready to begin to recover the usual formalism of quantum mechanics. The problems we are facing is that the composability two-product algebra looks nothing like the Hilbert space formulation, does not contain operator norm axioms, and has an unusual realization when $J^2 = +1$. 

We know that quantum and classical mechanics respect $J^2 = -1$ and $J^2 = 0$ respectively. We will start by investigating the $J^2 = +1$ case. We will look at a concrete realization of the composability two-product algebra in the case of hyperbolic composability (hyperbolic quantum mechanics) and attempt to eliminate it using physical arguments. We  will show that hyperbolic composability violates positivity and this can lead to overall negative probability predictions or ``ghosts''. Then we will investigate the phase space formalism of classical and quantum mechanics. We will not assume any mathematical structures and will derive the Poisson bracket for classical mechanics. For quantum mechanics we will derive a K\"ahler manifold which will be used to arrive at the usual Hilbert space formulation by deformation quantization. 

\subsection{Elimination of the hyperbolic composability solution} 
\label{EliminateHyperbolic}

Inspired by the phase space formulation of quantum mechanics (elliptic composability) we can introduce the phase space realization of hyperbolic quantum mechanics by using the following products:

\begin{eqnarray*}
\alpha &=& \frac{2}{\hbar}\sinh ( \frac{\hbar}{2}\overleftrightarrow{\nabla}) ,\\ \nonumber
\sigma &=& \cosh (\frac{\hbar}{2}\overleftrightarrow{\nabla}) .
\end{eqnarray*}

Similarly we can introduce a hyperbolic star product as well:

\begin{equation*}
f \star_{h} g = f \sigma g + \frac{J\hbar}{2} f \alpha g = f e^{\frac{J\hbar}{2} \overleftrightarrow{\nabla}} g ,
\end{equation*}

\noindent with $J^2 = +1$.

It is straightforward to check that those products satisfy all the relations of the composability two-product algebra.  

What is the hyperbolic star multiplication? This is nothing but a split-complex number multiplication for phase space functions defined over split-complex numbers. If complex numbers are defined as the Clifford algebra $Cl_{0,1}$, split complex numbers are defined as the Clifford algebra $Cl_{1,0}$ \cite{KhrennikovSegre}. Instead of $i=\sqrt{-1}$, split complex numbers have an imaginary unit $j=\sqrt{+1}$ with $j \neq 1$. For the phase space realization the simplest form of the composability parameter $J$ has the following form in matrix representation:

\begin{eqnarray*}
i = J &= \begin{pmatrix} 0 & -1 \\ 1 & 0 \end{pmatrix} &{\rm ~elliptic} ,\\
J &= \begin{pmatrix} 0 & 1 \\ 0 & 0 \end{pmatrix}  &{\rm ~parabolic} ,\\
j = J &= \begin{pmatrix} 0 & 1 \\ 1 & 0 \end{pmatrix}  &{\rm ~hyperbolic} .
\end{eqnarray*}

In complex quantum mechanics in phase space formulation, the expectation value of real star squares $g^{*}(x,p) \star g(x,p)$ is always positive even when the probability distribution contains negative parts: $\left< g^* \star g \right> \geq 0$.

The computation is as follows \cite{TCurtright}:
\begin{eqnarray*}
\int dxdp~ (g^* \star g) F = (2 \pi \hbar) \int dxdp~(g^* \star g) (F \star F) \\
= (2 \pi \hbar) \int dxdp~(g^* \star g) \star (F \star F) \\ 
= (2 \pi \hbar) \int dxdp~ (g^* \star g \star F ) \star F \\ 
= (2 \pi \hbar) \int dxdp~ F \star (g^* \star g \star F ) \\ 
= (2 \pi \hbar) \int dxdp~ (F \star g^* ) \star (g \star F ) \\ 
= (2 \pi \hbar) \int dxdp~ (F \star g^* ) (g \star F ) \\ 
= (2 \pi \hbar) \int dxdp~ {(g \star F)}^* (g \star F ) \\ 
= (2 \pi \hbar) \int dxdp~ {|F \star g |}^2 ,
\end{eqnarray*}

\noindent where $F$ is a not necessarily positive Wigner function \cite{WignerFunctions} corresponding to a pure state ($F = (2 \pi \hbar) F \star F$).

The same computation holds in hyperbolic quantum mechanics as well:\\ $\left< g^{*}\star_h g\right> = (2 \pi \hbar) \int dxdp~ {|F \star_h g |}^2 $. We observe that the final answer is given as an integral of a number of the form $z^* z$. In complex numbers this is always positive, but not in split-complex numbers and hence the hyperbolic composability theory contains unphysical negative probabilities. Next we want to confirm this finding by investigating the Hilbert space-like realization for hyperbolic composability and better understand the role of complex numbers in quantum mechanics by looking at how split-complex numbers affect the hyperbolic counterpart.

\subsection{Hilbert space-like realization for hyperbolic composability}
\label{HilbertHyperbolic}

It is helpful to understand what kind of Hilbert space-like formulation the hyperbolic case might have. It turns out that the usual functional analysis has a rich hyperbolic counterpart and it all starts from a reversed triangle inequality in some suitable generalization of the concept of norm. 

We will only present a high level overview without proofs of the new functional analysis domain, because we only need the results for their heuristic value. Additional details are presented in Appendix A. We will see that the Gelfand-Naimark-Segal (GNS) construction \cite{GNSReference} is not categorical in nature and therefore should not be attempted right away in the quantum mechanics reconstruction program.  

Both complex and split-complex numbers have polar decompositions. In the split-complex case the phase part is based on hyperbolic sines and cosines instead of the regular sines and cosines. In the split-complex case, the radius part of the decomposition is zero on the bisectors between the real and imaginary axis and the zero radius separates the hyperbolic complex plane in four quadrants. If you do not cross the quadrant boundaries, a reverse triangle inequality holds in each of the quadrants and this generates in turn an entire new functional analysis mathematical landscape. We will name the mathematical structures in this landscape the same way as their ``elliptical'' counterparts, but we will prefix them with ``para''.

There is a conversion dictionary between the usual functional analysis spaces and proofs and their corresponding hyperbolic counterparts:

\begin{table}[H] 
\caption{Conversion dictionary from regular functional analysis to hyperbolic functional analysis}
\label{ConversionDic}       
\begin{tabular}{ll}
\hline\noalign{\smallskip}
{\em Elliptic} & {\em Hyperbolic} \\
\noalign{\smallskip}\hline\noalign{\smallskip}
triangle inequality & reversed triangle inequality\\
sup & inf \\
convergent & divergent\\
bounded & unbounded\\
complete & incomplete\\
\hline\noalign{\smallskip}
\end{tabular}
\end{table}

As such we have para-Cauchy sequences, para-incompleteness, para-metric spaces, para-inner product spaces, and para-Hilbert spaces (a para-Hilbert space is a para-inner product space which is para-incomplete). The indefinite para-norm of a linear operator acting on a vector space over split-complex numbers can be defined as follows:

\begin{equation*}\label{operator norm}
||T|| = \inf_{\substack{ x\in \mathcal{D} (T)\\ ||x|| \ne 0}} \bigg| \frac{||T x||}{||x||}\bigg| {\rm sign} (||Tx||/||x||) .
\end{equation*}
 
We observe that the condition $||x|| \ne 0$ automatically prevents crossing the boundaries of the domain of the validity of the reversed triangle inequality. For split-complex numbers their indefinite para-seminorm $||z||$ is  defined as follows: 

\begin{equation*}\label{indefinite split complex seminorm}
|| z || = {\rm sign~}(z^*z) \sqrt{|z^* z|} .
\end{equation*}

In a vector space over split-complex numbers, the complex conjugation defines an involution, just like in their complex number counterparts. Because of this, a {\em Polarization Identity} holds as an algebraic identity:

\begin{eqnarray*}
x^* y = \frac{1}{4}[{(x+y)}^* (x+y) - {(x-y)}^* (x-y)] +\\ \nonumber
\frac{j}{4}[{(x+jy)}^* (x+jy) - {(x-jy)}^* (x-jy)] .
\end{eqnarray*}

Also the {\em Parallelogram Identity} holds as well:

\begin{equation*}
{(x+y)}^* (x+y) + {(x-y)}^* (x-y)= 2( x^* x + y^* y ) .
\end{equation*}

In turn this allows us to introduce an indefinite inner product as follows:

\begin{eqnarray*}\label{inner product definition}
<x, y> = x^* y  = \\ \nonumber
\frac{1}{4} [{(x+y)}^* (x+y) - {(x-y)}^* (x-y)] + \\ \nonumber
\frac{j}{4} [{(x+jy)}^* (x+jy) - {(x-jy)}^* (x-jy)]=\\ \nonumber
\frac{1}{4}[{\rm sign} (||x+y||) {||x+y||}^2 - {\rm sign} (||x-y||) {||x-y||}^2] + \\ \nonumber
\frac{j}{4}[{\rm sign} (||x+jy||) {||x+jy||}^2 - {\rm sign} (||x-jy||) {||x-jy||}^2] .
\end{eqnarray*}

State spaces demand considering convex sets regardless of composability classes. A key result in the elliptic composability case is that given a point $x$ in an inner product space $X$ and a complete not empty convex set $M$, there is a unique point $y \in M$ such that $||x-y||$ is minimal. This result is a prerequisite for subsequent important results like the factorization of Hilbert spaces in orthogonal complements, and for the Riesz representation theorem \cite{KreiszigBook}. 

This result does not hold in hyperbolic functional analysis and this prevents orthogonal decompositions for para-Hilbert spaces and a generalization of Riesz representation theorem. Hence in the hyperbolic case a GNS construction \cite{GNSReference} is untenable. More important, this shows that this construction is not a consequence of composability (categorical) arguments and we should not attempt proving directly the C*-algebra condition:

\begin{equation*}
||x^* x|| = {||x||}^2 .
\end{equation*}  

Instead we will need to find a different route proving the existence of the Hilbert space formulation. 

\subsection{Deriving the Poisson bracket and reconstructing classical mechanics}
\label{DerivePoisson}

In this section we assume that the collection of all Hamiltonian vector fields at a point $p$ span the tangent space $T_p(Q)$.

From the composability two-product algebra properties, in parabolic composability the the product $\sigma$ is commutative and associative. Hence it is isomorphic with regular function multiplication $f\sigma g = fg$. The product $\alpha$ can be proven to be a bracket as follows: 

We start with the simpler setting of an affine Poisson variety and consider the (affine) space $\mathbb{F}$ of polynomial functions on the cotangent bundle $M$. Assuming that the dimension of the configuration space $Q$ is $n$, we can define a bracket $\{ \cdot , \cdot \}$ on $\mathbb{F}[x_1, \dots , x_{2n}]$ in the cotangent bundle in the following way:

\begin{equation*}
\{ F , G \} = \sum_{i,j = 1}^{2n} \{ x_i, x_j \} \frac{\partial F}{\partial x_i} \frac{\partial G}{\partial x_j} ,
\end{equation*}

\noindent with $F, G \in \mathbb{F}[x_1, \dots , x_{2n}]$.

This is the most general way to construct a product $\alpha$ (which is a biderivation) and the proof is by induction using the argument that two biderivations of some commutative associative algebra $\mathcal{A} = \mathbb{F}$ are equal as soon as they agree on a system of generators for $\mathcal{A}$ \cite{PoissonBook}. If the product $\sigma$ is not associative (like in elliptic composability) this argument does not apply.

The reason the set $\{x_i \}$ contains twice as many elements as the dimension of $Q$ is that the cotangent bundle $M$ is itself a manifold of dimension $2n$. We started part two of the reconstruction project in the tangent plane and in general there is no natural way to identify vectors with co-vectors on a manifold. However in our case we have a vector field on $M$ which induces a vector field on $Q$ which can be thought as a function acting on the cotangent bundle. A vector field $X_q$ in $T_q Q$ is expressed in local coordinates as:
\begin{equation*}
X_q = \sum_{i} X^i (q) \frac{\partial}{ \partial q^i} .
\end{equation*}

The conjugate momentum map $P_X (q,p) = p(X_q )$ from the cotangent bundle $T^* Q$  to $\RR$ is defined for all cotangent vectors $p \in T^* Q$ and has the following expression:
\begin{equation*}
P_X (q,p) = \sum_i X^i (q) p_i ,
\end{equation*}

\noindent where $p_i$ is defined as the momentum function corresponding to the tangent vector $\partial /\partial q^i$:
\begin{equation*}
p_i = P_{\partial /\partial q^i}.
\end{equation*}

Therefore $q^i$ and $p_i$ form a coordinate system on the cotangent bundle $M$ and we can rename without loss of generality $x_{i}$ to $q^i$ and $x_{i+n}$ to $p_i$ for $i \in [1,n]$. 

Another way to see that the product $\alpha$ is the bracket from above is to recall that $\alpha$ was defined as the Lie derivative:

\begin{equation*}
f \alpha g = \lim_{\epsilon \rightarrow 0} \frac{g - T_{f_\epsilon} g}{\epsilon} ,
\end{equation*}

\noindent and this is identical with the bracket up to a normalization factor. 

We already proved that $\alpha$ is skew-symmetric and all that remains to be shown is that: $\{q^i, q^j \} = \{p_i, p_j \} = \{q^i, p_j \} = 0$ for $i \neq j$ and $\{q^i, p_i \} = -\{p_i, q^i \} = 1$. The proof follows from the bipartite fundamental relationship for $\alpha$:

\begin{equation*}
(q^i\otimes 1) \alpha_{ij} (1 \otimes q^j) = \{ q^i, q^j\} = (q^i\alpha 1)\otimes (1 \sigma q^j) + (q^i\sigma 1)\otimes (1 \alpha q^j) = 0 ,
\end{equation*}

\noindent and similarly for $ \{p_i, p_j \}$ and $\{q^i, p_j \}$.

Please note that the argument above does not imply that $\{q^i, p_i \} = 0$ because $q^i$ and $p_i$ belong to the same (sub)system:

\begin{equation*}
(q^i\otimes 1) \alpha_{ii} (p_i \otimes 1) = \{ q^i, p_i\} = (q^i\alpha p_i)\otimes (1 \sigma 1) + (q^i\sigma p_i)\otimes (1 \alpha 1) = \{ q^i, p_i\} .
\end{equation*}

If we normalize $\alpha$ such that $\{q^i, p_i \} = 1$, from the skew-symmetry we have $\{p_i, q^i \} = -1$, $\{q^i, q^i \} = \{p_i, p_i \} = 0$ and we recover the usual Poisson bracket:

\begin{equation*}
f \alpha g = \{ f , g \} = f \overleftrightarrow{\nabla} g = \sum_{i = 1}^{n}  \frac{\partial f}{\partial q^i} \frac{\partial g}{\partial p_i} - \frac{\partial f}{\partial p_i} \frac{\partial g}{\partial q^i} .
\end{equation*}

In the case of symplectic manifolds we can appeal to Darboux theorem \cite{PoissonBook} and obtain the same result as above. 

Because $f \sigma g = fg$, $M$ is now equipped with a Poisson algebra and is upgraded to a symplectic manifold. Hamilton's equations follow from the Lie derivative:

\begin{equation*}
\dot{g} = -f \alpha g = -\{ f, g\} = \{ g, f\} ,
\end{equation*} 
\noindent or in a more familiar form: $\dot{g}= \{g,H\}$ with $H = f$ and we recover the Hamiltonian formulation of classical mechanics.

\subsection{Symplectic vs. Poisson manifolds}
\label{SymplecticvsPoisson}

In the prior section we started with a nondegeneracy property which led to a symplectic manifold. We also made the assumption that $ \{ x_i, x_j \}$ is constant. When we no longer require nondegeneracy we generalize the symplectic manifold to a Poisson manifold which is a large topic in symplectic reduction \cite{LandsmanBook}, \cite{ButterfieldPaper}. The typical example is given by a free pivoted rigid body whose motion is described by the Euler's equations. This can be put in Hamiltonian form in an odd-dimensional Poisson manifold using a Lie-Poisson bracket. 

For Poisson varieties the most general Poisson bracket is still:

\begin{equation*}
\{ F , G \} = \sum_{i,j = 1}^{d} \{ x_i, x_j \} \frac{\partial F}{\partial x_i} \frac{\partial G}{\partial x_j} ,
\end{equation*}

\noindent with $F, G \in \mathbb{F}[x_1, \dots , x_{d}]$ but $ \{ x_i, x_j \}$ is no longer a constant.

Some non-symplectic Poisson manifolds can be obtained by reduction from a symplectic manifold using a Lie algebra and this corresponds to an alternative realization of a constrained dynamical system. 

For a general Poisson manifold, Darboux theorem states \cite{PoissonBook} that when the rank is locally constant and equal with $2r$ that there exists a coordinate neighborhood with coordinates $(q^1, \dots, q^r, p_1, \dots, p_r, z_1, \dots, z_s )$ such that:

\begin{equation*}
\{ F , G \} =  \sum_{i = 1}^{r}  \frac{\partial F}{\partial q^i} \frac{\partial G}{\partial p_i} - \frac{\partial F}{\partial p_i} \frac{\partial G}{\partial q^i} ,
\end{equation*}

\noindent and we are still able to define the Poisson bracket if we ignore the $z$ coordinates.

\subsection{Time evolution and composability classes}
\label{TimeEvolution}

To recover the elliptic composability products in the phase space formalism we will follow a deformation argument. Consider the fundamental relationships:

\begin{eqnarray*}
\alpha_{12} = \alpha_1 \sigma_2 + \sigma_1 \alpha_2 ,\\
\sigma_{12} = \sigma_1 \sigma_2 + \frac{J^2 \hbar^2}{4} \alpha_1 \alpha_2 , 
\end{eqnarray*}

\noindent where we explicitly revert back from the convention $\hbar^2 = 4$ to be able to use $\hbar$ as a free parameter which can be taken to zero.

From the prior section we know that when $\hbar = 0$ we have:

\begin{eqnarray*}
\alpha_0 = \alpha |_{\hbar = 0} &=& \overleftrightarrow{\nabla} ,\\
\sigma_0 = \sigma |_{\hbar = 0} &=& I ,\\
\end{eqnarray*}

\noindent where $I$ is the identity. Let us now demand that in general we have $\alpha$ and $\sigma$ functions of the Poisson bracket $\overleftrightarrow{\nabla}$:

\begin{eqnarray*}
\alpha = \cal{F} (\hbar \overleftrightarrow{\nabla}, \hbar) ,\\
\sigma = \cal{G} (\hbar \overleftrightarrow{\nabla}, \hbar) .\\
\end{eqnarray*}

One possible solution (which will later see that it only applies in flat space) is suggested by the formal analogy of the fundamental bipartite relationships with the polar decomposition of complex numbers:

\begin{eqnarray*}
\alpha &=& \frac{2}{\hbar}\sin (\frac{\hbar}{2}\overleftrightarrow{\nabla}) ,\\ \nonumber
\sigma &=& \cos (\frac{\hbar}{2}\overleftrightarrow{\nabla}) ,
\end{eqnarray*}

\noindent and those are the Moyal and the cosine brackets. 

We have seen that the product $\alpha_0 = \overleftrightarrow{\nabla}$ is unique because it is the Lie derivative in $M$ and we arrive at it from the invariance of the laws of nature under time evolution. What meaning can we attach to the elliptic composability product $\alpha$? Does invariance of the laws of nature under time evolution imply both $\alpha_0$ and $\alpha$? Both products satisfy the first three properties of the composability two-product algebra: Lie, Jacobi, and Leibniz. However in the elliptic case the invariance of the laws of nature under time evolution must preserve a non-trivial compatibility relationship as well. Excluding considerations of bi-Hamiltonian systems in general time evolution is defined by the product $\alpha$. In quantum mechanics $\alpha$ is no longer the Lie derivative (Moyal bracket is not the Poisson bracket), and instead of the Hamilton's equations of motion we have the Schr\"odinger equation which corresponds to a different kind of time evolution. In the next section we will see that quantum mechanics can be understood as constrained classical mechanics which preserves the non-trivial compatibility relation. 

The product $\alpha_0$ exists regardless of composability class and the deformation approach is mathematically well defined. The reconstruction of the composability two-product algebra proof using invariance under tensor composition arguments is still valid if we assume the existence of a product $\alpha$ which satisfies the Leibniz identity. Because $\alpha_0$ always exists we seek all possible deformations of $\alpha_0$ which preserve the Leibniz identity and there are two such deformation classes possible corresponding to the elliptic and hyperbolic cases.

\subsection{Deriving the phase space formulation for quantum mechanics}
\label{DerivingPhaseSpace}

In the prior section we derived the Moyal bracket and the cosine bracket, but how can we use them to make experimental predictions? Experiments consist of preparation followed by measurement and experiments can be composed \cite{GoyalQM}. Suppose we serially combine three experiments $A, B, C$ (this line of thought leads to the path integral formulation of quantum mechanics.). The outcome of the final measurement is independent of how we partition the experiments: $(AB)C, A(BC), ABC$ and this demands associativity. To perform computations related to experimental predictions we therefore need an associative product $\star$ build from the $\alpha$ and $\sigma$ products. We have seen such an example earlier in the form of Weyl-Groenewold (or star) product: $\star = \beta = \sigma + \frac{J \hbar}{2} \alpha$. 

Before investigating possible generalizations of the Weyl-Groenewold star product, let us consider the inverse problem. From an $\alpha$ and $\sigma$ product we can construct an associative product $\sigma + \frac{J \hbar}{2} \alpha$. Under what conditions can we reverse the operation and extract $\alpha$ and $\sigma$ from an associative product? Because the start product $\star$ is not commutative (the order of doing subsequent experiments matter), we can extract its symmetric and skew-symmetric parts as follows:

\begin{eqnarray*}
A \alpha^{'} B = \frac{J}{\hbar} (A \star B - B \star A) ,\\
A \sigma^{'} B = \frac{1}{2} (A \star B + B \star A) .\\
\end{eqnarray*}

Here we recognize the similarity with the commutator and the Jordan product from Hilbert space formulation. Direct computation shows that $\alpha^{'}$ and $\sigma^{'}$ obey the Leibniz, Jacobi, Jordan, compatibility, and the fundamental composability relations. To be fully equivalent with a composability two-product algebra we need the three additional properties: $\alpha^{'}$ to respect the relational property of the dynamic: $1 \alpha^{'} f = f\alpha^{'} 1 = 0$, $\sigma^{'}$ to be unital, and the star product to respect the $J$ involution. Relationality for $\alpha^{'}$ demands: $A\star 1 = 1 \star A$ which with unitality for $\sigma^{'}$: $A \sigma^{'} 1 = A$ demands unitality for $\star$: $A\star 1 = 1\star A = A$ .

Therefore for a noncommutative start product $\star$ to be equivalent with a composability two-product algebra, in the deformation approach we need three properties:

\begin{enumerate}
\item Associativity,
\item Unitality,
\item Compatibility with complex conjugation.
\end{enumerate} 

The most general form for the star product is:

\begin{equation*}
f\star g=\sum_{n} \hbar^n C_n(f,g) ,
\end{equation*}

\noindent where $C_n(f,g)$ is bidifferential operator of order $n$ subject to constraints generated by the three properties.

We already constructed such a product, the Weyl-Groenewold star product. In phase space formulation of quantum mechanics Wigner's functions \cite{WignerFunctions} are quasi-probabilities and to get to physical predictions we need to integrate them. The integration step however introduces considerations of convergence and the Weyl-Groenewold star product works only for flat manifolds, but Poisson manifolds in classical mechanics are not required to be flat, and quantum mechanics should not demand flatness either. Can we {\em always} construct an associative product for any Poisson manifold in deformation quantization? The answer is yes and was proven by Maxim Kontsevich in 1997 \cite{KontsevichPaper}.  

This very important result proves rigorously the existence of deformation quantization for standard quantum mechanics and from it we can always extract the composability two-product algebra. Therefore the phase space formulation of complex quantum mechanics is rigorously established. What we now seek is to pass from the phase space formulation to the Hilbert space formulation. This is equivalent to the transition from commutative to noncommutative geometry \cite{ConnesBook}.

\subsection{Deriving the K\"ahler manifold for quantum mechanics}
\label{DerivingKahler}

After deriving the Moyal and cosine brackets for the elliptic composability case, we now seek to understand the origin of the inner product in the Hilbert space formulation of quantum mechanics.

We start in flat $\RR^{2n}$ space and explicitly build a K\"ahler manifold. Then this is generalized to the case when we start from a non-flat space symplectic manifold.

\subsubsection{The flat space case}
\label{FlatCase}
 
Again we follow the deformation approach and we will analyze the elliptic case using the tools of parabolic composability. From the Poisson bracket used in the definition of the Moyal bracket we extract a symplectic form $\omega^{IJ}$. Let us call its inverse $\Omega_{IJ}$: $\omega^{IK} \Omega_{KJ} = \delta^I_J$. Please note that in this section we are following the convention of reference \cite{BengtssonBook}. In elliptic composability we have a parameter $J$ satisfying $J^2 = -1$. Because the Hamiltonian formalism is defined over the real numbers, $J$ cannot be a scalar and must have a matrix representation which we now attempt to construct. To simplify the problem we consider a one-dimensional physical system. In this case the maximum matrix dimension for the representation of $J$ is two (there are only two coordinates in the cotangent bundle: $p$ and $q$ ) and it is larger than one ($J$ cannot be a scalar when we assume the number system to be $\RR$). The only possibility is for $J$ to have the same representation as the representation of the complex numbers imaginary unit: 

\begin{equation*}
i = J = 
\begin{pmatrix} 
0 & -1 \\ 1 & 0 
\end{pmatrix} .
\end{equation*}

The definition is up to an overall sign (which defines the complex conjugation involution property of quantum mechanics). We see that $J$ performs a swap of $q$ and $p$ and this easily generalizes to the n-dimensional case. In general, $J$ is not the imaginary complex number unit but a tensor of rank $(1,1)$: $J = J^I_{~J}$ with the following matrix representation:  

\begin{equation*}
J^I_{~J} = \begin{bmatrix} 0 & -1_n \\ 1_n & 0 \end{bmatrix} .
\end{equation*}

$J$ is an ``almost-complex'' structure because is defined on $M$ at each tangent space. Invariance of the laws of nature under time evolution demands that $J^I_{~J}$ and  $\Omega_{IK}$ to be preserved under time evolution and therefore we can construct a metric tensor $g$ preserved under time evolution as well as follows: $g_{IJ}=\Omega_{IK} J^K_{~J}$ with:

\begin{eqnarray*}
\Omega_{IJ} =& \begin{bmatrix} 0 & 1_n \\ -1_n & 0 \end{bmatrix} ,\\
g_{IJ} =& \begin{bmatrix} 1_n & 0 \\ 0 & 1_n \end{bmatrix} .\\
\end{eqnarray*}

By construction we have $g = \Omega J$ and by inspection we see that $\Omega = J^T g$ where $J^T$ is the transpose of $J$. Also the metric tensor $g$ defines a Hermitean structure because it satisfies: $g(JX, JY ) = g(X, Y)$:

\begin{eqnarray*}
{(J^I_{~K}X^K)}^T g_{IJ} (J^J_{~P}Y^P) = {(X^K)}^T {(J^I_{~K})}^T {\Omega}_{IQ} J^Q_{~J} J^J_{~P} Y^P = \\
{(X^K)}^T {(J^I_{~K})}^T (-{\Omega}_{IQ} \delta^Q_P ) Y^P = {(X^K)}^T {(J^I_{~K})}^T {(\Omega_{PI})}^T Y^P = \\
{(X^K)}^T {(\Omega_{PI} J^I_{~K})}^T Y^P = {(X^K)}^T {(g_{PK} )}^T Y^P = {(X^K)}^T g_{KP} Y^P ,
\end{eqnarray*}

\noindent where $T$ defines matrix or vector transposition. A complex inner product can now be introduced as $g + i \Omega$:

\begin{equation*}
< X , Y > = X^T g Y + i X^T \Omega Y ,
\end{equation*}

\noindent where:

\begin{equation*}
X = \begin{pmatrix} 
q^1 \\
\vdots \\
q^n\\
p_1\\ 
\vdots \\
p_n\\
\end{pmatrix} .\\
\end{equation*}

We therefore constructed an almost complex manifold for the elliptic composability case. The almost complex manifold is integrable when the Nijenhuis tensor $N$ \cite{NijenhuisTensor}:

\begin{equation*}
N(R, S ) = [R, S] + J[JR, S ] + J[R, JS ] - [JR, JS ] ,
\end{equation*} 

\noindent defined on vector fields $R$ and $S$ vanishes. When this happens the almost complex manifold becomes a K\"ahler manifold because $\omega$ is closed by invariance under flow lines. 

But what does it mean that an almost complex manifold is not integrable? Given any point $p\in M$, it is not possible to find coordinates such that J takes the canonical form from above on an entire neighborhood of $p$ and hence $J^2 \ne -1$. This means that we are breaking the fundamental composability relations and the description of the laws of nature is no longer invariant under arbitrary tensor composition. Therefore we must have a K\"ahler manifold.

Time evolution preserves $\Omega$ and $J$ by preserving a normalization constraint:

\begin{equation*}
\langle g \rangle - 1 = X^I g_{IJ} X^J = 0 .
\end{equation*}

The constrained Hamilton's equations of motion give rise to the Schr\"odinger equation and demands that the observables are Hermitean (equivalently they commute with $J$ because time evolution cannot change the composability class) \cite{BengtssonBook}. 

\subsubsection{The non-flat space case}
\label{NonFlatSpace}

Passing from the flat to non-flat case, we have to simply replace the Moyal sine bracket with: $J/\hbar (A \star B - B \star A)$ where $\star$ is no longer defined over flat space, but over a general symplectic manifold. We can still extract the symplectic form $\omega$, and we still have the $(1,1)$ tensor $J$, but now we loose their explicit representation. However, all arguments from above still apply and we found ourselves into a general K\"ahler manifold setting from which we would need to extract operators on a Hilbert space.     

\subsection{Berezin quantization and C*-algebras}
\label{BerezinQuantization}

To complete the derivation of complex quantum mechanics we will follow a deformation quantization approach able to extract operators in a Hilbert space. 

To this aim we will use Berezin quantization because K\"ahler manifolds often admit a Berezin quantization and Berezin quantization enjoys the advantage of positivity \cite{LandsmanBook}. Depending on the physical system under consideration, other approaches are possible like Weyl quantization \cite{WeylQuantization} for flat spaces. Berezin quantization is actually a dequantization but we will not enter into technical details \cite{GiachettaBook}.

A K\"ahler manifold is not quantizable in general unless we have a ``quantum line bundle'': $\{ L, h, \nabla \}$ where $L$ is a holomorphic line bundle, $h$ is a hermitean metric, and $\nabla$ is a connection satisfying a compatibility condition. We have seen that Poisson manifolds always admit a deformation quantization \cite{KontsevichPaper} and hence the K\"ahler manifold obtained in the prior section must also be quantizable provided there is a relationship between the Kontsevich star product and Berezin star product. However the relationship between Berezin and Kontsevich quantization is nontrivial \cite{KazunoriWakatsuki}.    

For compact K\"ahler manifolds we can prove the existence of the quantum line bundle. We have used the composability arguments in constructing the K\"ahler manifold, and the proof of quantization cannot come from them. But we have not yet used positivity. A K\"ahler manifold is quantizable when the line bundle $L$ is positive and this allows the use of Kodaira embedding theorem \cite{KodeiraThm}: ``If $L$ is a line bundle on a compact complex manifold, then $L$ is ample if and only if $L$ is positive.'' \cite{HodgeBook}.

Then if $L$ is ample the K\"ahler manifold can be embedded in a complex projective space \cite{Schlichenmaier}. 

On $\RR^{2n}$ the Berezin quantization \cite{BerezinQuantization} is the following prescription to construct compact operators from continuous functions on phase space:

\begin{equation*}
Q_{\hbar} (f) = \int_{\RR^{2n}} \frac{dp dq}{2 \pi \hbar} f(p,q) |\Phi_{\hbar}^{(p,q)} \rangle \langle \Phi_{\hbar}^{(p,q)} | ,
\end{equation*}

\noindent where $\Phi_{\hbar}^{(p,q)}$ are coherent states defined as:

\begin{equation*}
\Phi_{\hbar}^{(p,q)} = {(\pi \hbar)}^{-1/4} e^{-ipq/2 \hbar} e^{ipx/\hbar} e^{-{(x-q)}^2 /2 \hbar} .
\end{equation*}

At this point we recovered the Hilbert space formulation of complex quantum mechanics. This can be double checked by using the GNS construction \cite{GNSReference} after extracting the C*-algebra condition for any bounded operators $T$ as follows:

\begin{eqnarray*}
 {||T \Phi||}^2 = \langle T \Phi, T \Phi \rangle =  \langle T^* T \Phi , \Phi \rangle \leq ||T^* T \Phi ||~||\Phi|| \leq ||T^* T||~||\Phi||^2\\
{||T||}^2 \leq ||T^* T|| \leq ||T^*|| ||T|| = {||T||}^2\\
||T^* T|| = {||T||}^2 .
\end{eqnarray*}  

We have seen that recovering quantum mechanics formalism (even for infinite dimensional Hilbert spaces) consists of two steps: a categorical derivation of the composability two-product algebra followed by a non-functorial quantization step whose existence is guaranteed by positivity. But why is this last step non-functorial?

The reason is that the star product can be understood as an infinite sum of terms proportional with the powers of the Planck constant $\hbar$. Also the star product being associative by construction, associativity transfers to each term in all Planck constant power terms. Then a natural question to ask is the equivalence of two star products. The equivalence classes of star products on symplectic manifolds are in one-to-one correspondence with second de Rham cohomology $H^2_{dR}(M)$. Therefore there could be inequivalent ways of quantization.

\subsection{Distinguishing quantum from classical mechanics}
\label{DistinguishingQuantum}

It is time to collect all the results so far and complete the quantum mechanics reconstruction program. From classical and quantum mechanics a composability two-product algebra was extracted. Then this mathematical structure was derived from physical principles using very general categorical arguments. It was found that the invariance of the laws of nature under tensor composition admits three ``fixed points'': elliptic, parabolic, and hyperbolic composability. The hyperbolic domain generalizes regular functional analysis and positivity is not possible. Hence this solution is unphysical.

For elliptic and parabolic composability we obtained the usual Hamiltonian formalism, which in the elliptic case has an additional structure of a metric space which gives rise to a K\"ahler manifold. This implies the existence of an inner product. The standard complex number quantum mechanics is recovered using Berezin quantization which is only one of the possible ways to add positivity and the norm axioms into the composability two-product algebra formalism.  

We note that there could not be any consistent mixed classical-quantum description of a physical system because a quantum system cannot have any back-reaction on a classical system \cite{SahooPlanck} and the composability classes are disjoint. Therefore there is only one last step needed to distinguish between classical and quantum mechanics. 

The only distinguishing property in the composability two-product algebra was the parameter $J$ which for classical mechanics respects $J^2 = 0$, and for quantum mechanics respects $J^2 = -1$. $J^2 = -1$ is responsible for quantum superposition. In turn this shows that quantum mechanics violates Bell's inequalities \cite{Bell1} and nature confirms the violation \cite{AspectExperiment}. While there are still experimental loopholes waiting to be closed, there is no single experiment to date which contradicts quantum mechanics. 

\section{Discussions}
\label{Discussions}

\subsection{The number system for quantum mechanics and transition spaces}
\label{NumberSystem}

When we add the positivity condition to the composability two-product algebra we impose a reality condition on the Jordan algebra $\sigma$: $A^2 + B^2 = 0 \Rightarrow A=B=0$ making them ``real Jordan algebras''. Their full classification is well known \cite{JordanAlgebras} and is related with projective spaces over the division algebras. 

The number system for a representation of quantum mechanics is defined by the representation of the simplest possible quantum state with only one degree of freedom. In this case the dynamic correspondence map $J$ is the same as the imaginary unit of complex numbers, and the natural number system for quantum mechanics are the complex numbers. In turn this leads to the concept of a transition probability space \cite{LandsmanBook}. 

How are we to understand then quantum mechanics over quaternions or real numbers? A comprehensive monograph on quaternionic quantum mechanics was written by Adler \cite{AdlerQuaternions} and quaternionic quantum mechanics can be understood as a constrained complex quantum mechanics system \cite{BengtssonBook}. Real quantum mechanics is defined over a number system which is too small to accommodate dynamic correspondence and it has to be embedded in complex numbers \cite{AdlerQuaternions}. We note that the actual wavefunction in real or quaternionic quantum mechanics is different than their complex counterpart because the inner product is different \cite{AdlerQuaternions}. Still, their predictions are not in any way distinct than the predictions of complex quantum mechanics in the common range of validity. 

Because the only division number systems possible are: real numbers, complex numbers, quaternions, and octonions, and because octonionic quantum mechanics is forbidden ($\sigma$ must be the real part of an associative involutive product), it looks that in the current framework there is no other possibility for a quantum mechanics number system. This is an incorrect conclusion however because there is a hidden unnecessary assumption in the argument: the transition space must be a transition {\em probability} space. In other words, we require Born's interpretation and Born's rule.

What if whenever we repeat an experiment, the end result is not a probability (a real number), but we get something with a richer mathematical structure instead? At first this seems absurd, but what if we get a {\em probability 4-current}? A remarkable recent result obtained by Grgin \cite{GrginBook} shows that there is a fourth number system possible which is related to a spin factor Jordan algebra. The new number system is called a ``quantion'' and was so named because of its similarity with quaternions. 

A quantion has the following matrix representation:
\begin{equation*}
q= \begin{pmatrix} 
q_1 & q_3 & 0 & 0 \\ q_2 & q_4 & 0 & 0 \\ 0 & 0 & q_1 & q_3 \\ 0 & 0 & q_2 & q_4 
\end{pmatrix} ,
\end{equation*}

\noindent with $q_1, q_2, q_3, q_4 \in \CC$.

Quantionic quantum mechanics generalizes the Born rule, reduces itself to the complex quantum mechanics in the non-relativistic limit, demands Dirac's equation, is equivalent with Dirac's spinors, has the $CPT$ discrete symmetry, and corresponds to quantionic projective space instead of $\CC P^n$. If $q$ is a quantion, it has a polar decomposition into a future oriented 4-vector and a $U(1) \times SU(2)$ phase term. $q^{\dagger}q$ is a future oriented 4-vector corresponding to a current probability density respecting a relativistic continuity equation \cite{GrginBook}. In fact $q^{\dagger}q$ is the Dirac current. Quantionic quantum mechanics is inherently relativistic and while it does not make new physical predictions because is equivalent with Dirac's spinors (quantions correspond to a different factorization of the d'Alembertian), their composability two-product algebra realization corresponds to a constrained quantum system over $SO(2,4)$ understood as a ring. $SO(2,4)$ was selected as a non $SU(n)$ realization of quantum mechanics starting from Cartan's classification of Lie algebras when one demands the additional requirement of the composition two-product algebra relation of compatibility \cite{foundationPaper1,foundationPaper2,foundationPaper3,foundationPaper4}.

Suppose that the number system for quantum mechanics is selected to be $SO(2,4)$. Because $SO(2,4) \sim SU(2,2)$, it is not possible to implement positivity directly. However, the negative probabilities or ``ghosts'' can be eliminated if we pick a particular element of the composability two-product algebra to play the role of $J$ and demand that all observables in the $SO(2,4)$ quantum mechanics commute with that element. What results is a composability sub-two-product algebra which is equivalent with a composability two-product algebra over quantions. Quantionic quantum mechanics corresponds to C*-Hilbert modules instead of C*-algebras and this shows first that not all quantum mechanics realizations are corresponding to C*-algebras, and second that adding positivity is a non-trivial problem even when the number of degrees of freedom remains finite. Additional details are presented in Appendix B.

\subsection{The collapse postulate and the measurement problem}
\label{CollapsePostulate}

For real and complex quantum mechanics, Born rule follows from Gleason's theorem \cite{GleasonThm} and quantum mechanics is inherently probabilistic. In between measurements the time evolution is unitary, but after an experiment is performed and an experimental outcome is recorded, the wavefunction collapses. How are we to understand the collapse postulate? Is it just an update of information, a rederivation of a Hilbert space representation of a quantum system? 

We seek to derive the collapse postulate from a pure unitary time evolution using categorical arguments. This does not imply that the many-worlds interpretation \cite{Vaidman1} is a mathematical necessity nor that the epistemic point of view is invalid. We will show that the collapse postulate is forced upon us by ignoring a mathematical structure not unlike in the early days of special relativity people talked about ``imaginary $ict$'' time because they ignored the metric tensor.
 
Here is how we proceed. Recall that we proved that $\mathcal{U}(M, \otimes, \mathbb{R}, \alpha, \sigma, J)$ is a commutative monoid. This encodes the idea that two physical systems can be considered together and that their description is invariant under tensor composition. However, it can also represent a description of two {\em interacting} systems. The collapse postulate can be understood as an inverse operation to the tensor product. Can we upgrade the composition class $\mathcal{U}$ from a commutative monoid to a group? Such a construction is known as the Grothendick group construction \cite{GrothendieckGroup} and is pure categorical. All we need to proceed is an equivalence relationship. Do we have a natural equivalence relationship in quantum mechanics? The answer is yes and it is a consequence of elliptic composability. The Grothendick group construction is not possible in classical mechanics (parabolic composability). 

The equivalence relationship comes from a swap symmetry: what the quantum system can unitarily evolve over here can be undone by another unitary evolution of the environment over there. In other words in quantum mechanics we encounter envariance \cite{ZurekEnvariance}.

Let us formally define the equivalence relationship. We call two pairs of a Cartesian product of wavefunctions equivalent:

\begin{equation*}
( {\left| \psi \right>}_p , {\left| \psi \right>}_n) \sim ({\left| \phi \right>}_p , {\left| \phi \right>}_n) ,
\end{equation*}

\noindent if given any unitary transformation $U_p$ acting on the left element $(\left| \psi \right>, \cdot )$ there exists a unitary transformation $U_n$ acting on the right element $(\cdot, \left| \psi \right> )$, a wavefunction $\left| \xi \right>$, and a unitary transformation $U_{\xi}$ such that:

\begin{equation*} 
{\left| \psi \right>}_p \otimes {\left| \phi \right>}_n \otimes \left| \xi \right> = 
(U_{p} {\left| \phi \right>}_p ) \otimes (U_n {\left| \psi \right>}_n) \otimes (U_{\xi} \left| \xi \right>) .
\end{equation*}

The ancilla $\left| \xi \right>$ is required by the proof of the transitivity property and can be ignored when we want to prove the reflexivity and symmetry properties.

The left element of the Cartesian product is called the ``positive'' element (the system), while the right one is called the ``negative'' element (the measurement device and the environment). 

The proof of the equivalence properties (symmetry, reflexivity, and transitivity) is presented in Appendix C. 

Obtaining the Grothendieck group of composability is only the first step in solving the measurement problem because an element of the Grothendieck group is an equivalence class containing all possible experimental outcomes. To explain why there is only one outcome we need a mechanism to spontaneously break the Grothendieck equivalence class. 

Let us make two observations: an experimental outcome contains many copies of the outcome information and the system and measurement device sometimes are in an unstable equilibrium. Infinitesimal perturbations can exponentially grow and lead to a unique peak of the wavefunction \cite{LandsmanFlea}. Although this mechanism was proven exactly in a few cases, more is required to research this approach and establish its universality. Its relationship with the quantum Darwinism program \cite{ZurekDarwin} has to be researched as well. If we want to stay outside the usual quantum mechanics interpretations, as of now the measurement problem is an open problem. 

\subsection{Quantum mechanics and relativity}
\label{QMRelativity}

The current approach of deriving the Hilbert space of quantum mechanics is very similar with how Lorentz transformations can be derived in the special theory of relativity. 

Special theory of relativity is rooted in two simple postulates: the laws of nature are invariant under changes in inertial frames of reference and the principle of invariant light speed. The first postulate of special theory of relativity tells us that there is no absolute reference frame and there are no distinguished speeds (except for the speed of light). The second postulate is ultimately justified by experimental evidence. 

If one starts with the invariance of consecutive space-time transformations under inertial reference frame transformations (in addition with the symmetries of space and time) one obtains Lorentz and Galilean transformations. Then nature selects which transformation occurs. 

\begin{figure}[h]
\includegraphics[width=0.35\textwidth]{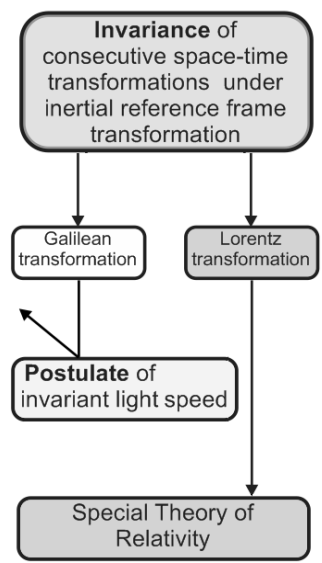}
\caption{Deriving special theory of relativity line of argument.}
\label{fig:LineOfArgumentSTR}
\end{figure}

The derivation of quantum mechanics follows a similar pattern. First we have the invariance of the laws of nature under time evolution and under tensor composition. To them we add a relational postulate similar in spirit with the absence of absolute reference frame. From this we derive three solutions. One of them is eliminated by positivity which demands the ability to define a state able to make predictions (probabilistic or deterministic) about nature. The final solution is selected based on experimental evidence as well. 

\begin{figure}[h]
\includegraphics[width=0.5\textwidth]{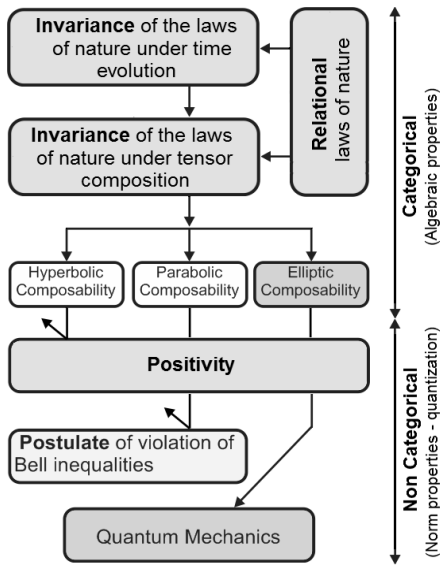}
\caption{Deriving quantum mechanics line of argument.}
\label{fig:LineOfArgumentQM}
\end{figure}

However there is tension between the two theories because quantum correlations between spatially separated regions cannot be causally explained, and yet quantum mechanics cannot be used to send signals faster than the speed of light. Currently there is no consensus on how to understand quantum mechanics in relation to locality. Is quantum mechanics local but not realistic, or is quantum mechanics non-local? (The disagreement is mostly on the meaning of the terms realism and locality. Bell locality is violated by nature but this is not an universally accepted definition of locality.) From the current derivation we see that quantum mechanics is locality-blind because considerations of distance do not enter the derivation in any form.  

In the fundamental relationship of the bipartite product $\sigma_{12}$, the lack of factorizability using only $\sigma_1$ and $\sigma_2$ is the root cause of the tension with the mathematical properties of space (compare this with classical mechanics and its compatibility with the idea of spatial separated regions). If we take the point of view of space-time and demand a causal explanation for quantum correlations, we are implicitly taking the point of view of parabolic composability and demanding an explanation of elliptic composability in the parabolic framework. 

Just like in special theory of relativity one does not attempt to explain Lorentz transformations using unbounded speed (Galilean) models, in quantum mechanics the real mystery is not its interpretation using a parabolic composability paradigm, but the emergence of space-time in a quantum mechanics framework.

\subsection{Comparison with other quantum mechanics reconstruction approaches}
\label{Comparison}

It is informative to compare the current quantum mechanics reconstruction approach with other reconstruction programs. While there is some degree of overlap, we can classify reconstruction approaches in six broad classes: based on the correspondence principle, based on path integral formulation, based on observables, based on an instrumentalist approach, based on quantum information, and based on quantum logic.

\subsubsection{Quantum mechanics reconstruction approaches based on the correspondence principle}
\label{CorrespondencePrinciple}

The current approach belongs into this class and is an extension of the original Grgin-Petersen \cite{GrginCompPaper} approach which was recently formalized in category formalism by Kapustin \cite{AKapustin}. Deformation quantization \cite{Bayen} also belongs in the correspondence principle class of approaches and we have shown that the composability two-product algebra based on the products $\alpha$ and $\sigma$ is equivalent with the associative and noncommutative star product approach in deformation quantization: $\alpha$ and $\sigma$ can define the star product, and the star product can define $\alpha$ and $\sigma$. 

Segal's approach \cite{SegalAxioms} is also inspired by a common classical-quantum mechanics set of axioms. But unlike the observables-generators duality Segal chooses an observables-states duality. His approach is more restrictive than quantum mechanics on phase space because it includes norm axioms (under his classification ``metric postulates'') and is also more general because his states product (which generalizes the start product $\star$) does not need to admit a skew-symmetric product. If we follow Rob Clifton's approach and restrict Segal algebras to what he calls ``Segalgebras'' (real closed linear subspaces of observables that contain the identity and are closed under
the (generally nonassociative) symmetric and antisymmetric products) we simply get self-adjoint parts of subalgebras of C*-algebras \cite{CliftonSegalgebras}.

\subsubsection{Quantum mechanics reconstruction approaches based on the path integral formulation}
\label{PathIntegral}

Goyal, Knuth, and Skilling approach \cite{GoyalQM} belongs in a category of its own and is similar in spirit with the current approach. They consider performing experiments sequentially or in parallel and extract natural symmetries obeyed by those experiment compositions. In turn, with an additional postulate inspired by Bohr's principle of complementarity, this introduces constraints which allow deriving explicit realization for probabilities. Then the rule for complex number multiplication in Feynman's path integral formulation is obtained. 

However, now we can recognize that their result is actually a derivation of the fundamental composability relation.

\subsubsection{Quantum mechanics reconstruction approaches based on observables}
\label{Observables}

Quantum mechanics can be defined over a phase or Hilbert space. In phase space one can take the operational approach of positive-operator valued measures (POVMs) \cite{SchroeckBook}. The standard GNS construction \cite{GNSReference} employs the associative product, but one can also start from the Jordan algebra of observables. In this case one can use the Jordan-GNS construction of Alfsen, Shultz, and St{\o}rmer \cite{JordanGNS}.  

Barnum and Wilce derived quantum mechanics from three axioms: (1) individual systems are Jordan algebras, (2) composites are locally tomographic, and (3)  at least one system has the structure of a qubit \cite{BarnumWilce}. They also characterize finite dimensional quantum theory among probabilistic theories having the structure of a dagger-monoidal category which is related with Kapustin's categorical formulation \cite{AKapustin}.

\subsubsection{Quantum mechanics reconstruction approaches based on an instrumentalist approach}
\label{InstrumentalistApproach}

Lucien Hardy \cite{HardyQM} introduced an instrumentalist approach based on five ``reasonable'' axioms. He starts by introducing two numbers: the number of degrees of freedom $K$ (the number of real parameters required to specify the state) and the dimension N (the maximum number of states that can be reliably distinguished from one another in a single shot measurement).

With those two definitions, the axioms are: (1) Probabilities, (2) Simplicity ($K$ is determined by a function of $N$ and for each given $N$, $K$ takes the minimum value consistent with the axioms), (3) Subspaces, (4) Composite systems rules ($N_{A\otimes B} = N_A N_B$ and $K_{A\otimes B} = K_A K_B$), (5) Continuity (there exists a continuous reversible transformation on a system between any two pure states of that system).

Then Hardy derives a key property: $K = N^r$ which recovers classical mechanics for $r=1$ and complex quantum mechanics for $r = 2$. His equation requires a slight generalization in the quantionic quantum mechanics case: $K = 4N^2$ because there are four spinor components in Dirac's equation.

Hardy's approach was further refined by Jochen Rau \cite{RauPaper} in a Bayesian paradigm. He extended Hardy's dimensionality arguments and computed the impact of preparation, composition, and continuity on the structure group which is proven to be the unitary group.

\subsubsection{Quantum mechanics reconstruction approaches based on quantum information}
\label{QuantumInformation}

John Bell proved that quantum mechanics correlations (which can be expressed using operator norm) cannot be attributed to shared randomness (local hidden variables) \cite{Bell1}. However those correlations cannot be used to transmit signals faster than the speed of light, and Popescu and Rohrlich flipped the problem around and asked if nonlocality can be used as an axiom for quantum theory \cite{PR1}. This led to the introduction of hypothetical Popescu-Rohrlich boxes which can achieve super-quantum correlations. Then there are three levels of correlations possible: hidden variable correlations corresponding to classical mechanics, quantum correlations, and super-quantum correlations and this started the search for information theoretical criteria to distinguish them. Since composability demands only classical and quantum mechanics, PR boxes cannot occur in nature if the laws of nature are invariant under tensor composition.

However, quantum information approaches proved fruitful and Rob Clifton, Jeffrey Bub, and Hans Halvorson \cite{BubPaper} working in the formalism of C*-algebras showed that observables and the state space of a physical theory are quantum mechanical if one starts from three information theoretical axioms: (1) the impossibility of superluminal information transfer between two physical systems by performing measurements on one of them; (2) the impossibility of perfectly broadcasting the information contained in an unknown physical state; (3) the impossibility of unconditionally secure bit commitment.

Quantum mechanics reconstruction can be achieved using information theoretical considerations. Dakic and Brukner \cite{DakicBrukner} introduced four axioms: (1) (Information capacity) an elementary system has the information carrying capacity of at most one bit. All systems of the same information carrying capacity are equivalent; (2) (Locality) the state of a composite system is completely determined by local measurements on its subsystems and their correlations; (3) (Reversibility) between any two pure states there exists a reversible transformation; (4) (Continuity) between any two pure states there exists a continuous reversible transformation.

Lluis Masanes and Markus M\"uller \cite{MasanesQM} considered five axioms: (1) in systems that carry one bit of information, each state is characterized by a finite set of outcome probabilities; (2) the state of a composite system is characterized by the statistics of measurements on the individual components; (3) all systems that effectively carry the same amount of information have equivalent state spaces; (4) any pure state of a system can be reversibly transformed into any other; (5) in systems that carry one bit of information, all mathematically well-defined measurements are allowed by the theory. When continuity is imposed on axiom five one recovers quantum theory.

Chiribella, D'Ariano, and Perinotti \cite{ChiribellaDerivation} derived quantum mechanics from five axioms: (1) Causality: the probability of a measurement outcome at a certain time does not depend on the choice of measurements that will be performed later. (2) Perfect distinguishability: if a state is not completely mixed (i.e. if it cannot be obtained as a mixture from any other state), then there exists at least one state that can be perfectly distinguished from it, (3) Ideal compression: every source of information can be encoded in a suitable physical system in a lossless and maximally efficient fashion. Here lossless means that the information can be decoded without errors and maximally efficient means that every state of the encoding system represents a state in the information source, (4) Local distinguishability: if two states of a composite system are different, then we can distinguish between them from the statistics of local measurements on the component systems, (5) Pure conditioning: if a pure state of system AB undergoes an atomic measurement on system A, then each outcome of the measurement induces a pure state on system B. (Here atomic measurement means a measurement that cannot be obtained as a coarse-graining of another measurement). Quantum mechanics is singled out by a purification principle: ``Every state has a purification. For fixed purifying system, every two purifications of the same state are connected by a reversible transformation on the purifying system''.

Barnum, M\"uller, and Ududec \cite{BarnumMullerUdudec} introduced four axioms: (1) Classical Decomposability: every state of a physical system can be represented as a probabilistic mixture of perfectly distinguishable states of maximal knowledge (``pure states''); (2) Strong Symmetry: every set of perfectly distinguishable pure states of a given size can be reversibly transformed to any other such set of the same size; (3) No Higher-Order Interference: the interference pattern between mutually exclusive ``paths'' in an experiment is exactly the sum of the patterns which would be observed in all two-path subexperiments, corrected for overlaps; (4) Observability of Energy: there is non-trivial continuous reversible time evolution, and the generator of every such evolution can be associated to an observable (``energy'') which is a conserved quantity.

We note that in all quantum information approaches Hardy's continuity axiom is present in one form or another and this singles out quantum mechanics. Although Barnum, M\"uller, and Ududec's approach avoid talking about composite systems, their ``Observability of Energy'' postulate is related to ``dynamic correspondence'' between observables and generators which is a consequence of a invariance under tensor composition.

\subsubsection{Quantum mechanics reconstruction approaches based on quantum logic}
\label{QuantumLogic}

This approach is the oldest of all reconstruction approaches and is represented by Piron's result \cite{PironQM}. This approach is related with projective spaces and departure from Boolean algebras signals non-classical behavior. However, non-Boolean algebras do not necessarily imply quantum mechanics because there could be classical macroscopic systems which violate non-Boolean algebras for appropriate definitions of propositions \cite{AertsBell}.

\subsubsection{Composition and information role in the foundation of quantum mechanics}
\label{CompositionInformation}

Both composition and information considerations are essential in deriving quantum mechanics. On one hand, pure composition arguments are unable to eliminate non-physical theories like hyperbolic quantum mechanics. On the other hand, all information theoretical axiomatic systems from above cannot avoid talking about composite systems. The role of composition arguments is unsurprising given Bell's theorem because for a single particle there are classical models which reproduce exactly all quantum mechanics predictions. 

Composition arguments are special cases of categorical considerations which we proved that determine the algebraic relationships. Information theoretical considerations demand positivity as a pre-requisite, and positivity is responsible for the norm properties in Hilbert space formulation. Positivity is very general and does not demand Born interpretation as quantionic quantum mechanics counterexample shows. Positivity only requires that ``{\em it} is what can generate a bit'', but the mathematical realization of the bit is left unspecified (bit, qubit, current probability density).

We have seen that the distinction between quantum and classical mechanics has a composition (algebraic), not an information origin. We appeal to experimental evidence in the form of Bell's inequalities to distinguish between classical and quantum mechanics, but any ``quantumness'' principle can work, like Hardy's instrumentalist continuity axiom, or Feynman principle that in quantum mechanics probability amplitudes rather than probabilities superimpose \cite{AdlerQuaternions}.

Why does nature prefer elliptic composition over parabolic composition? This no different from asking why nature prefers having a maximum speed limit over unlimited speeds. For quantum mechanics the answer cannot be given by either categorical or positivity arguments and it is very likely that the answer will only come from a self-contained ``theory of everything''. Maybe only an universe with elliptic composition is able to create itself.

\subsection{Open problems}
\label{OpenProblems}

Quantum mechanics derivation started with the invariance of the laws of nature under time evolution. As such it seems that the concept of time is a primitive notion and not dependent on quantum mechanics. However, Tomita-Takesaki theory shows the existence of a distinguished one-dimensional parameter which can be understood as time and time seems to have a noncommutative origin \cite{ConnesBook}. This point of view was expanded into a ``thermal time hypothesis'' \cite{ThermalTime} and now we may face a problem of circularity:  quantum mechanics is derived by assuming time, but is time a consequence of quantum mechanics as well? 

The composability two-product algebra has two classes of realizations: commutative phase spaces and non-commutative Hilbert spaces. The domain of the Hilbert space realizations appears in general to be smaller because while in phase space the functions are continuous, in Hilbert space realizations they are analytic (all complex holomorphic functions are analytic), and there are continuous functions which are not analytic. But does this lead to physically meaningful differences? 

So far we have not discussed the infinite degrees of freedom case. There the Stone-von Neumann \cite{StoneTheorem} theorem does not hold and there are infinite classes of unitarily inequivalent Hilbert space representations. Also Hilbert spaces are not available in the interaction picture \cite{HaagTheorem} despite the practical success of the perturbation approach of quantum field theory. The infinite number of unitarily inequivalent Hilbert space representations should be considered in an instrumentalist approach which takes into account the smearing effects of physical devices used in actual measurements. The position of this paper is that of ``algebraic imperialism'' in that the physical content is encoded in algebraic properties, and the Hilbert space realization are only a computation convenience of no ontological significance \cite{HalvorsonPaper}. At this time we do not have a complete classification of all possible Hilbert space realizations of quantum mechanics. What we do have is a complete classification only for finite degrees of freedom for transition probability spaces. 

The complete classification for the allowed number systems of quantum mechanics is another open problem. The existence of four number systems shows that while the inner product of kets and bras is universal, Born interpretation is not and depends on the concrete number system of the realization. However, all quantum mechanics realizations: phase space, Hilbert spaces over real, complex, quaternionic, and quantionic numbers give identical predictions when they share a domain of validity. 

\section*{Appendix A}
\label{appendix1}

\subsection*{Split-complex numbers preliminaries}
\label{SplitComplex}

We denote split-complex numbers as $\DD$. Like complex numbers, split complex numbers have an ``imaginary unit'' $j \ne 1$ but with a different property: $j^2 = +1$. This makes $\DD$ an involutive algebra.

If $x$ and $y$ are the real and imaginary components of a split-complex number $z= x + j y$, there are four possible hyperbolic polar form decompositions based on the values of $x$ and $y$:

\begin{eqnarray*}
z = +\rho (\cosh \theta + j \sinh \theta) &{\rm ~if~} x>0 {\rm ~and~} |x| > |y| ,\\ \nonumber
z = +\rho (\sinh \theta + j \cosh \theta) &{\rm ~if~} y>0 {\rm ~and~} |y| > |x| ,\\ \nonumber
z = -\rho (\cosh \theta + j \sinh \theta) &{\rm ~if~} x<0 {\rm ~and~} |x| > |y| ,\\ \nonumber
z = -\rho (\sinh \theta + j \cosh \theta) &{\rm ~if~} y<0 {\rm ~and~} |y| > |x| ,\\ \nonumber
\end{eqnarray*}

\noindent with $\rho$ a positive real number and $\theta$ a real number. We can call $\rho$ the modulus and $\theta$ the hyperbolic phase. Because we can define an (indefinite) inner product in split-complex quantum mechanics, just like in its complex quantum mechanics counterpart, {\em pure states are defined only up to a hyperbolic phase}. Standard functional analysis spaces can be introduced and used, but they lack a physical interpretation. Instead we need to consider spaces respecting the state definition. This requires introducing a new functional analysis domain. The starting point is investigating the triangle inequality which is part of the metric space definition and leads to considerations of convergence, completness, and norms.

Split-complex numbers do not respect a triangle inequality, but using the polar form decompositions we see that inside each of the four areas separated by the zero norm boundaries (defined as $\rho =0$) a reversed triangle inequality holds:

\begin{equation*}
\bigg| || z + w || \bigg| \geq \bigg| || z || \bigg| + \bigg| || w || \bigg| .
\end{equation*}

\noindent The root cause of triangle inequality for complex numbers is the fact that the cosine function is bounded from above, while the root cause of the reversed triangle inequality is the fact that the hyperbolic cosine function is bounded from below.

\subsection*{Para-metric and para-normed spaces}
\label{ParaMetric}

{\bf Definition~}{\it A semi para-metric space is a pair $(X,d)$, where $X$ is a set and $d$ is a semi para-distance function on $X$ defined on $X \times X$ such that for all $x,y,z \in X$ we have:

\begin{eqnarray*}
d\in \RR_{+}~~~~~(PM1) ,\label{(PM1)}\\ 
d(x, y) = 0 {\rm ~if~} x=y~~~~~(PM2) ,\label{(PM2)}\\ 
d(x,y)=d(y,x)~~~~~(PM3) ,\label{(PM3)} \\ 
d(x,y) \geq d(x, z) + d(z, y)~~~~~(PM4) ,\label{(PM4)} \\
\end{eqnarray*}

\noindent with $x,y,z$ in $(PM4)$ connectable by a path not crossing any zero distances points.
}

An example of para-metric space is $\DD$ itself.

{\bf Definition~}{\it A sequence $\left\{ x_n \right\}$ in a metric space $X = (X,d)$ is said to be para-Cauchy if for every $\epsilon > 0$ there is an $N = N(\epsilon)$ such that:
\begin{equation*}
d(x_m , x_n) > \epsilon {\rm ~for~every~ } m,n > N .
\end{equation*}
The space $X$ is said to be para-incomplete if every para-Cauchy sequence in $X$ diverges.
}

To build an intuition about para-Cauchy sequences and para-incompleteness, when the reversed triangle inequality holds one thinks not of (elliptic) boundary value problems, but of (hyperbolic) initial value problems and preservation of causality. Physically it is desirable to shield the local value by the influence of far away points when the topology is non-Hausdorff. 

{\bf Definition~}{\it An indefinite para seminormed space is a vector space $X$ over split complex numbers $\DD$ with a (not necessarily positive) real-valued function ${||x||}_{X}$ for all $x \in X$ obeying the following properties:
\begin{eqnarray*}
{||\alpha x||}_{X} = {||\alpha||}_{\DD} {||x||}_{X} ~~~~~(PN1) ,\label{(PN1)}\\
\bigg| ||x+y|| \bigg| \geq \bigg| ||x|| \bigg| + \bigg| ||y|| \bigg| ~~~~~(PN2) ,\label{(PN2)}
\end{eqnarray*}

\noindent where $\alpha \in \DD$ and with $x,y$ in $(PN2)$ connectable by a path not crossing any zero (para) norm points.
}

\noindent Given any two para-normed spaces, we can consider linear maps between them. In particular, functionals are linear maps to $\DD$ which leads us to the concept of dual spaces. Unlike regular functional analysis, the interesting cases here are the unbounded maps defined as follows:

{\bf Definiton~}{\it Let $X$ and $Y$ be para-normed spaces and $T:\mathcal{D} (T) \rightarrow Y$ a linear operator, where $\mathcal{D}(T) \subset X$. The operator $T$ is said to be unbounded if there is a positive real number $c$ such that for all $x\in \mathcal {D}(T)$
\begin{equation*}
\bigg| ||Tx|| \bigg| \geq c \bigg| ||x|| \bigg| .
\end{equation*}
}

We can also define the corresponding norm of linear operators with the difference that we are this is defined as infimum and not as supremum:

{\bf Definition~}{\it The number $||T||$ defined as:
\begin{equation*}
||T|| = \inf_{\substack{ x\in \mathcal{D} (T)\\ ||x|| \ne 0}} \bigg| \frac{||T x||}{||x||}\bigg| {\rm sign} (||Tx||/||x||) ,
\end{equation*}

\noindent is called the para-norm of operator $T$.
}

\noindent If we consider the algebraic properties in addition to norm properties we can introduce a para-normed algebra as follows:

{\bf Definition~}{\it A para-normed algebra $A$ in a para-normed space which is an algebra such that for all $x,y\in A$:
\begin{equation*}
\bigg| ||x y|| \bigg| \geq \bigg| ||x|| \bigg| \bigg| ||y|| \bigg| ,
\end{equation*}

\noindent when $||x||*||y|| > 0$.
}

Considering para-Cauchy behavior we can restrict the prior definition:

{\bf Definition~}{\it A para-Banach algebra $A$ is a para-normed algebra which is para-incomplete.}

The following theorem regarding para-normed algebras holds:
 
{\bf Theorem~}{\it The linear operator algebra between two para-normed spaces is a para-normed algebra.
}

\subsection*{Para-inner product spaces}
\label{ParainnerProduct}

From the indefinite inner product we can prove:

{\bf Theorem~}{\it If $x, y \in \DD$ and $||x|| * ||y|| \geq 0$, the following para-Cauchy-Schwarz inequality holds:
\begin{equation*}
|<x,y>| \geq ||x|| ||y|| .
\end{equation*}
}

{\bf Definition~}{\it A para-Hilbert space is an indefinite para-inner product space which is para-incomplete.}

\subsection*{No-go result for orthogonal decomposition and Riesz representation}

The following result prevents the existence of orthogonal decompositions for para-Hilbert spaces and generalization of Riesz representation theorem:

{\bf Theorem~}{\it Suppose $X$ is an indefinite inner product space over $\DD$ and $M\ne \varnothing$ a complete convex subset. If $\forall x \in X, \exists y \in M$ such that $\delta = \inf_{\bar{y} \in M} || x - \bar{y} || = ||x-y||$, then $y$ is not necessarily unique.
}
\begin{proof}
Suppose there is $y_0 \in M$ such that $||x-y|| = ||x-y_0 || = \delta > 0$. Then ${||y-y_0||}^2 = {||(y-x) -(y_0 -x)||}^2$ and by the parallelogram identity:
\begin{eqnarray*}
{||y-y_0||}^2 {\rm sign}(||y-y_0||) = 2 {||y-x||}^2 + 2 {||y_0 - x||}^2\\
 - {||(y-x) + (y_0 - x)||}^2 {\rm sign}(||(y-x) + (y_0 - x)||) \\
=2 \delta^2 + 2\delta^2 -4 {||\frac{1}{2} (y+y_0) - x||}^2 {\rm sign}(||\frac{1}{2} (y+y_0) - x||)
\end{eqnarray*} 

Since $M$ is convex, $\frac{1}{2}(y+y_0) \in M$ and $||\frac{1}{2} (y+y_0) -x|| \geq \delta$. 

In turn this implies:

${||y-y_0||}^2 {\rm sign}(||y-y_0||) \leq 0$.

If the norm is positive definite this would imply that $y_0 = y$, but if the norm is indefinite the uniqueness is no longer a mathematical necessity. \qed 
\end{proof}

\section*{Appendix B}
\label{appendix2}

Here we present additional information about quantions and quantionic quantum mechanics. Quantionic quantum mechanics was discovered by Emile Grgin and its physical interpretation is due to Nikola Zovko. The results in this appendix are not new, but quantionic quantum mechanics is not well known. Quantionic quantum mechanics represents an important counterexample to the GNS construction in adding positivity to the quantum formalism, because quantions are not a division number system and is impossible to construct a non-trivial quotient space. Instead, quantionic quantum mechanics corresponds to a C*-Hilbert module.  

If $Q$ is a quantion defined as:

\begin{equation*}
Q= \begin{pmatrix} 
 a & c & 0 & 0 \\ b & d & 0 & 0 \\ 0 & 0 & a & c \\ 0 & 0 & b & d 
\end{pmatrix} ,
\end{equation*} 

\noindent with $a, b, c, d \in \CC$, a ``reduced quantion'' $q$ is defined as:

\begin{equation*}
q= \begin{pmatrix} 
a & c \\ b & d  
\end{pmatrix} .
\end{equation*} 

\subsection*{Two ``norms'' of quantions}
\label{TwoNorms}

For any quantion $Q$ one can define an ``algebraic norm'' $A(Q) = Q^{\dagger} Q$ and a ``metric norm'' $M(Q) = Q^{\sharp} Q = {\rm det~} q I = (ad-bc) I$ where:

\begin{eqnarray*}
Q^{\dagger} &=& \begin{pmatrix} 
 a^* & b^* & 0 & 0 \\ c^* & d^* & 0 & 0 \\ 0 & 0 & a^* & b^* \\ 0 & 0 & c^* & d^* 
\end{pmatrix} ,\\
Q^{\sharp} &=& \begin{pmatrix} 
 d & -c & 0 & 0 \\ -b & a & 0 & 0 \\ 0 & 0 & d & -c \\ 0 & 0 & -b & a 
\end{pmatrix} .
\end{eqnarray*} 

Here $Q^{\sharp}$ is called a ``metric dual''.

The main theorem is that the two norms commute: $AM(Q) = MA(Q)$. $A(Q)$ is a future-oriented four-vector, $M(Q)$ is a complex number, and $AM(Q)$ is a positive real number.

\subsection*{Quantion-spinor relation}
\label{QuantionSpinor}

If we introduce the column representation for a quantion Q:

\begin{equation*}
Q = \begin{pmatrix} 
a \\ b \\ c \\ d  
\end{pmatrix} ,
\end{equation*}

\noindent and consider the Dirac spinor $\Psi$:

\begin{equation*}
\Psi= \begin{pmatrix} 
\psi_1 \\ \psi_2 \\ \psi_3 \\ \psi_4  
\end{pmatrix} ,
\end{equation*}

\noindent the relationship between a spinor and a quantion is:

\begin{equation*}
\begin{pmatrix} 
\psi_1 \\ \psi_2 \\ \psi_3 \\ \psi_4  
\end{pmatrix} = \frac{1}{\sqrt{2}}
\begin{pmatrix} 
c \\ -a \\ b^* \\ d^*  
\end{pmatrix} .
\end{equation*}

The quantionic current $j^\mu = Q^{\dagger} Q$ is the same as Dirac's current: $j^\mu = \Psi^{\dagger} \gamma^0 \gamma^\mu \Psi$. The d'Alembertian $\square$ can be decomposed either in the usual way as: $\square = D^2$ with $D=\gamma^\mu \partial_\mu$, or as in the quantionic way as: $\square = {\cal{D}}^{\sharp} \cal{D}$ with:

\begin{equation*}
\cal{D} = \begin{pmatrix} 
D I_{2 \times 2} & \delta I_{2 \times 2} \\ \delta^* I_{2 \times 2} & \triangle I_{2 \times 2}  
\end{pmatrix} ,
\end{equation*}

\noindent where $D, \delta, \triangle$ are the Newman-Penrose symbols:

\begin{eqnarray*}
D &=& \partial_0 + \partial_3 , \\
\delta &=& \partial_1 + i \partial_2 , \\
\triangle &=& \partial_0 - \partial_3 .
\end{eqnarray*}

\subsection*{Quantionic quantum mechanics}
\label{QQM}

In quantionic quantum mechanics any spacetime point has associated a quantion and this defines a quantionic bundle. Nikola Zovko proposed to impose a quantionic four-vector continuity condition: $\partial_\mu (Q^{\dagger} Q) = 0$ which generalizes Born's rule and leads to Dirac's equation. Then the algebraic norm $A$ extracts a current probability density and corresponds to quantum mechanics, and the metric norm $M$ extracts the probabilities in a frame of reference and corresponds to special relativity. The metric dual $Q^{\sharp}$ defines a parity transformation $P$, the algebraic dual $Q^{\dagger}$ defines a charge transformation $C$, and together $CP$ defines a time reversal operation $T$. $Cq = q$ defines the Minkowski space, $Pq = q$ defines complex numbers (and quantionic quantum mechanics reduces to complex quantum mechanics), and $Tq = q$ defines real quaternions (and quantionic quantum mechanics reduces to quaternionic quantum mechanics).

\section*{Appendix C}
\label{Appendix3}

In this appendix we establish the usual properties of an equivalence relationship (reflexivity, symmetry, and transitivity) in case of swap symmetry. Let us recall the definition of the equivalence relationship. We call two pairs of a Cartesian product of wavefunctions equivalent:

\begin{equation}
( {\left| \psi \right>}_p , {\left| \psi \right>}_n) \sim ({\left| \phi \right>}_p , {\left| \phi \right>}_n)
\end{equation}

\noindent if given any unitary transformation $U_p$ acting on the left element $(\left| \psi \right>, \cdot )$ there exists a unitary transformation $U_n$ acting on the right element $(\cdot, \left| \psi \right> )$, a wavefunction $\left| \xi \right>$, and a unitary transformation $U_{\xi}$ such that:

\begin{equation} \label{eq:Def}
{\left| \psi \right>}_p \otimes {\left| \phi \right>}_n \otimes \left| \xi \right> = 
(U_{p} {\left| \phi \right>}_p ) \otimes (U_n {\left| \psi \right>}_n) \otimes (U_{\xi} \left| \xi \right>) .
\end{equation}

\subsection*{Reflexivity}
\label{Reflex}

To prove reflexivity we need to show that: $(\left| a \right> , \left| b \right>) \sim (\left| a \right> , \left| b \right>)$. This means that for any $U_p$, exists a $U_n$ such that 

\begin{equation}
\left| a \right> \otimes \left| b\right>  = 
(U_{p} \left| a \right> ) \otimes (U_n \left| b \right>) ,  
\end{equation}

\noindent and this is the original definition of envariance with the positive elements the system and the negative elements the environment. The proof is by Schmidt decomposition \cite{ZurekEnvariance}. For any: 

\begin{equation}
U_p = \sum_{k=1}^{N} e^{i \phi_k} \left| a_k \right> \left< a_k \right| ,
\end{equation}

\noindent we have:

\begin{equation}
U_n = \sum_{k=1}^{N} e^{-i (\phi_k + 2 \pi l_k)} \left| b_k \right> \left< b_k \right| ,
\end{equation}

\noindent where $\left| a\right> \otimes \left| b\right> = \sum_{i=1}^{N} \lambda_i \left| a_i\right> \left| b_i \right> $ and $l_k$ arbitrary natural numbers.

\subsection*{Symmetry}
\label{Symm}

Suppose that $(\left| a\right>, \left| b\right>) \sim (\left| c\right>, \left| d\right>)$. This means that given any $U_p$ there exists $U_n$ such that:
\begin{equation} \label{eq:S1}
\left| a\right> \otimes \left| d\right> = (U_p \left| c \right>) \otimes (U_n \left| b \right>) .
\end{equation}

To prove symmetry we need to show that  $(\left| c\right>, \left| d\right>) \sim (\left| a\right>, \left| b\right>)$ is true as well. Then given any $V_p$ there exists a $V_n$ such that: 
\begin{equation}
\left| c\right> \otimes \left| b\right> = (V_p \left| a \right>) \otimes (V_n \left| d \right>) . 
\end{equation}

Using Eq.~(\ref{eq:S1}):
\begin{equation}
(V_p \left| a \right>) \otimes (V_n \left| d \right>) = (V_p U_p \left| c \right>) \otimes (V_n U_n \left| b \right>) ,
\end{equation}

\noindent we want $V_p U_p = 1$ and $V_n U_n = 1$. Observing that $U_n = U_n (U_p)$, for any $V_p$, pick $U_p = V_p^{-1}$ and then $V_n(V_p ) = U_n^{-1} = U_n^{-1} (V_p^{-1})$. 

\subsection*{Transitivity}
\label{Trans}

For transitivity, we need to show that if $(\left| a \right> , \left| b \right>) \sim (\left| c \right> , \left| d \right>)$ and $(\left| c \right> , \left| d \right>) \sim (\left| e \right> , \left| f \right>)$ then $(\left| a \right> , \left| b \right>) \sim (\left| e \right> , \left| f \right>)$

From the first equivalence, given any $U_p$, there exists $U_n$, $\left| \xi \right>$, and $U_{\xi}$ such that:
\begin{equation}\label{eq:firstEquiv}
\left| a \right> \otimes \left| d \right> \otimes \left| \xi \right> = 
(U_p \left| c \right>) \otimes (U_n \left| b \right>) \otimes (U_{\xi} \left| \xi \right>) ,
\end{equation}

\noindent and that given any $V_p$, there exists $V_n$, $\left| \eta \right>$, and $V_{\eta}$ such that:
\begin{equation}\label{eq:secondEquiv}
\left| c \right> \otimes \left| f \right> \otimes \left| \eta \right> = 
(V_p \left| e \right>) \otimes (V_n \left| d \right>) \otimes (V_{\eta} \left| \eta \right>) .
\end{equation}

We need to show that given any $W_p$, there exists $W_n$, $\left| \chi \right>$, and $W_{\chi}$ such that:
\begin{equation}
\left| a \right> \otimes \left| f \right> \otimes \left| \chi \right> = 
(W_p \left| e \right>) \otimes (W_n \left| b \right>) \otimes (W_{\chi} \left| \chi \right>) .
\end{equation}

From Eqs.~(\ref{eq:firstEquiv}) and (\ref{eq:secondEquiv}), we have:
\begin{eqnarray}
\left| a \right> \otimes \left| f \right> \otimes (\left| d \right> \otimes \left| c \right> \otimes \left| \xi \right> \otimes \left| \eta \right>) =~~~~~~~~~~~~~~~~~~~~~~~ \\ \nonumber
(V_p \left| e \right> ) \otimes (U_n \left| b \right>) \otimes \left[ U_p \left| c \right> \otimes U_{\xi} \left| \xi \right> \otimes V_n \left| d \right> \otimes V_{\eta} \left| \eta \right>\right] .
\end{eqnarray}

Now given any $W_p$ pick $U_p = V_p = W_p$. Define $W_n = W_n (W_p) = U_n (U_p ) = U_n (W_p)$. Then we have:
\begin{equation}
\left| \chi \right> = \left| d \right> \otimes \left| c \right> \otimes \left| \xi \right> \otimes \left| \eta \right> ,
\end{equation} 
\noindent and
\begin{equation}
W_{\chi} \left| \chi \right> = 
 V_n \left| d \right> \otimes U_p \left| c \right> \otimes U_{\xi} \left| \xi \right> \otimes V_{\eta} \left| \eta \right> ,
\end{equation}

\noindent which defines the unitary transformation $W_{\chi}$:

\begin{equation}
W_{\chi} = V_n(W_p) \otimes W_p \otimes U_{\xi} \otimes V_{\eta} .
\end{equation}

This concludes the proof of transitivity and Eq.~(\ref{eq:Def}) defines an equivalence relationship which in turn allows us to construct the Grothendieck group of composability.

\begin{acknowledgements}
I want to thank Emile Grgin for extensive correspondence. I want to thank Franklin E. Schroeck Jr., Howard Barnum, Caslav Brukner, and Ovidiu-Cristinel Stoica for useful comments and correspondence.
\end{acknowledgements}

\bibliographystyle{spmpsci}      
\bibliography{QMReconstructionFoP}   

\begin{thebibliography}{10}
\providecommand{\url}[1]{{#1}}
\providecommand{\urlprefix}{URL }
\expandafter\ifx\csname urlstyle\endcsname\relax
  \providecommand{\doi}[1]{DOI~\discretionary{}{}{}#1}\else
  \providecommand{\doi}{DOI~\discretionary{}{}{}\begingroup
  \urlstyle{rm}\Url}\fi

\bibitem{AdlerQuaternions}
Adler, S.L.: {Quaternionic Quantum Mechanics and Quantum Fields}.
\newblock International Series of Monographs on Physics. Oxford Univ. Press,
  New York (1995)

\bibitem{AertsBell}
Aerts, D.: {Example of a macroscopical situation that violates Bell
  inequalities}.
\newblock Lett.\ Nuovo\ Cim. \textbf{34}(4), 107--111 (1982)

\bibitem{AlfsenShultz}
Alfsen, E.M., Shultz, F.W.: {Geometry of State Spaces of Operator Algebras}.
\newblock Mathematics: Theory and Applications. Birkh\"auser, Boston (2003)

\bibitem{JordanGNS}
Alfsen, E.M., Shultz, F.W., St{\o}rmer, E.: {A Gelfand-Neumark theorem for
  Jordan algebras}.
\newblock Adv.\ Math. \textbf{28}(1), 11--56 (1978)

\bibitem{GNSReference}
Arveson, W.: {An Invitation to C*--Algebra}.
\newblock Springer-Verlag, New York (1981)

\bibitem{AspectExperiment}
Aspect, A., Grangier, P., Roger, G.: {Experimental Realization of
  Einstein-Podolsky-Rosen-Bohm Gedankenexperiment: A New Violation of Bell's
  Inequalities}.
\newblock Phys.\ Rev.\ Lett. \textbf{49}(2), 91--94 (1982)

\bibitem{GrothendieckGroup}
Atiyah, M.F., Anderson, D.W.: {K-theory}.
\newblock No.~7 in Mathematics Lecture Notes. W. A. Benjamin, New York and
  Amsterdam (1967).
\newblock Lectures by Atiyah (Fall 1964), notes by Anderson. Russian
  translation published as \textit{Lekcii po} K-\textit{teorii} (1967). 2nd
  edition published in 1989. MR:0224083.

\bibitem{BarnumMullerUdudec}
Barnum, H., M\"{u}eller, M.P., Ududec, C.: {Higher-order interference and
  single-system postulates characterizing quantum theory}.
\newblock {arXiv:1403.4147}  (2014)

\bibitem{BarnumWilce}
Barnum, H., Wilce, A.: {Local tomography and the Jordan structure of quantum
  theory}.
\newblock {arXiv:1202.4513}  (2012)

\bibitem{Bayen}
Bayen, F., Flato, M., Fronsdal, C., Lichnerowicz, A., Sternheimer, D.:
  {Deformation theory and quantization. I. Deformations of symplectic
  structures}.
\newblock Annal.\ Phys. \textbf{111}(1), 61--110 (1978)

\bibitem{Bell1}
Bell, J.S.: {On the Einstein-Poldolsky-Rosen paradox}.
\newblock Physics \textbf{1}(3), 195--200 (1964)

\bibitem{BengtssonBook}
Bengtsson, I., Zyczkowski, K.: {Geometry of Quantum States}.
\newblock Cambridge Univ. Press, Cambridge (2006)

\bibitem{BerezinQuantization}
Berezin, F.A.: {General Concept of Quantization}.
\newblock {Comm.\ Math.\ Phys.} \textbf{40}(2), 153--174 (1975).
\newblock \urlprefix\url{http://projecteuclid.org/euclid.cmp/1103860463}

\bibitem{HodgeBook}
Bertin, J., Demailly, J.P., Illusie, L., Peters, C.: {Introduction to Hodge
  Theory}.
\newblock American Mathematical Society, Rhode Island (2002)

\bibitem{ButterfieldPaper}
Butterfield, J.: {On Symplectic Reduction in Classical Mechanics}.
\newblock In: J.~Butterfield, J.~Earman (eds.) {Philosophy of Physics Part A
  (Handbook of the Philosophy of Science)}, pp. 1--132. North Holland, The
  Netherlands (2007)

\bibitem{ChiribellaQM}
Chiribella, G., D'Ariano, G.M., Perinotti, P.: {Probabilistic theories with
  purification}.
\newblock Phys. Rev. A \textbf{81}(6), 062,348 (2010).
\newblock \urlprefix\url{http://arxiv.org/abs/0908.1583}

\bibitem{ChiribellaDerivation}
Chiribella, G., D'Ariano, G.M., Perinotti, P.: {Informational derivation of
  quantum theory}.
\newblock Phys. Rev. A \textbf{84}(1), 012,311 (2011).
\newblock \urlprefix\url{http://arxiv.org/abs/1011.6451}

\bibitem{CliftonSegalgebras}
Clifton, R.: {Beables in Algebraic Quantum Mechanics}.
\newblock arXiv:quant-ph/9711009  (1997)

\bibitem{BubPaper}
Clifton, R., Bub, J., Halvorson, H.: {Characterizing Quantum Theory in Terms of
  Information-Theoretic Constraints}.
\newblock Found.\ Phys. \textbf{33}(11), 1561--1591 (2003).
\newblock \urlprefix\url{http://arxiv.org/abs/quant-ph/0211089}

\bibitem{ArtinWedderburn}
Cohn, P.M.: {Basic Algebra: Groups, Rings, and Fields}.
\newblock Springer-Verlag, London (2003)

\bibitem{ConnesBook}
Connes, A.: {Noncommutative Geometry}.
\newblock Academic Press, San Diego (1994)

\bibitem{ThermalTime}
Connes, A., Rovelli, C.: {Von Neumann algebra automorphisms and
  time-thermodynamics relation in generally covariant quantum theories}.
\newblock Class.\ Quant.\ Grav. \textbf{11}(12), 2899 (1994).
\newblock \urlprefix\url{http://arxiv.org/abs/gr-qc/9406019}

\bibitem{TCurtright}
Curtright, T.L., Zachos, C.K.: {Quantum Mechanics in Phase Space}.
\newblock Asia\ Pac.\ Phys. Newslet. \textbf{01}(01), 37--46 (2012).
\newblock \urlprefix\url{http://arxiv.org/abs/1104.5269}

\bibitem{DakicBrukner}
Dakic, B., Brukner, C.: {Quantum Theory and Beyond: Is Entanglement Special?}
\newblock In: H.~Halvorson (ed.) {Deep Beauty: Understanding the Quantum World
  through Mathematical Innovation}, pp. 365--392. Cambridge University Press,
  Cambridge (2011).
\newblock \urlprefix\url{http://arxiv.org/abs/0911.0695}

\bibitem{EmchBook}
Emch, G.G.: {Algebraic Methods in Statistical Mechanics and Quantum Field
  Theory}.
\newblock Wiley-Interscience, New York (1972)

\bibitem{NijenhuisTensor}
Fr\"{o}licher, A., Nijenhuis, A.: {Theory of vector-valued differential forms
  Part I.}
\newblock Indagat.\ Math. \textbf{59}, 338--359 (1956)

\bibitem{FuchsQM}
Fuchs, C.A.: {Quantum Mechanics as Quantum Information (and only a little
  more)}.
\newblock {arXiv:quant-ph/0205039}  (2002)

\bibitem{GiachettaBook}
Giachetta, G., Mangiarotti, L., Sardanashvily, G.: {Geometric and Algebraic
  Topological Methods in Quantum Mechanics}.
\newblock World Scientific, Singapore (2005)

\bibitem{GleasonThm}
Gleason, A.: {Measures on the Closed Subspaces of a Hilbert Space}.
\newblock Indiana Univ. Math. J. \textbf{6}(4), 885--893 (1957)

\bibitem{GoyalQM}
Goyal, P., Knuth, K.H., Skilling, J.: {Origin of Complex Quantum Amplitudes and
  Feynman's Rules}.
\newblock Phys.\ Rev.\ A\ \textbf{81}(2), 022,109 (2010).
\newblock \urlprefix\url{http://arxiv.org/abs/0907.0909}

\bibitem{foundationPaper1}
Grgin, E.: {Inherently Relativistic Quantum Theory. Part I. The Algebra of
  Observables}.
\newblock Fizika B \textbf{10}(3), 113--138 (2001).
\newblock \urlprefix\url{http://fizika.hfd.hr/fizika\_b/bv01/b10p113.htm}

\bibitem{foundationPaper2}
Grgin, E.: {Inherently Relativistic Quantum Theory. Part II. Classifications of
  the Solutions}.
\newblock Fizika B \textbf{10}(3), 139--160 (2001).
\newblock \urlprefix\url{{http://fizika.hfd.hr/fizika\_b/bv01/b10p139.htm}}

\bibitem{foundationPaper3}
Grgin, E.: {Inherently Relativistic Quantum Theory. Part III. Quantionic
  Algebra}.
\newblock Fizika B \textbf{10}(4), 187--210 (2001).
\newblock \urlprefix\url{{http://fizika.hfd.hr/fizika\_b/bv01/b10p187.htm}}

\bibitem{foundationPaper4}
Grgin, E.: {Inherently Relativistic Quantum Theory. Part IV. Quantionic
  Theorems}.
\newblock Fizika B \textbf{10}(4), 211--234 (2001).
\newblock \urlprefix\url{{http://fizika.hfd.hr/fizika\_b/bv01/b10p211.htm}}

\bibitem{GrginBook}
Grgin, E.: {The Algebra of Quantions. A Unifying Number System for Quantum
  Mechanics and Relativity}.
\newblock Authorhouse, Indiana (2005)

\bibitem{GrginCompPaper}
Grgin, E., Petersen, A.: {Algebraic implications of composability of physical
  systems}.
\newblock {Comm.\ Math.\ Phys.} \textbf{50}(2), 177--188 (1976).
\newblock \urlprefix\url{http://projecteuclid.org/euclid.cmp/1103900192}

\bibitem{HaagTheorem}
Haag, R.: {On quantum field theories}.
\newblock Dan. Mat. Fys. Medd. \textbf{29}(12), 1--37 (1955)

\bibitem{HalvorsonPaper}
Halvorson, H., M\"ueger, M.: {Algebraic Quantum Field Theory}.
\newblock In: J.~Butterfield, J.~Earman (eds.) {Philosophy of Physics Part A
  (Handbook of the Philosophy of Science)}, pp. 732--922. North Holland, The
  Netherlands (2007)

\bibitem{HardyQM}
Hardy, L.: {Quantum theory from five reasonable axioms}.
\newblock arXiv:quant-ph/0101012  (2001)

\bibitem{JordanAlgebras}
Jordan, P., von Neumann, J., Wigner, E.: {On an Algebraic Generalization of the
  Quantum Mechanical Formalism}.
\newblock Ann.\ Math. \textbf{32}(1), 29--64 (1934)

\bibitem{AKapustin}
Kapustin, A.: {Is quantum mechanics exact?}
\newblock {J.\ Math.\ Phys.} \textbf{54}, 062,107 (2013).
\newblock \urlprefix\url{http://arxiv.org/abs/1303.6917}

\bibitem{KhrennikovSegre}
Khrennikov, A., Segre, G.: {Von Neumann Uniqueness Theorem doesn't hold in
  Hyperbolic Quantum Mechanics}.
\newblock {Int.\ J.\ Theor.\ Phys.} \textbf{45}(10), 1869--1890 (2006).
\newblock \urlprefix\url{http://arxiv.org/abs/math-ph/0511044}

\bibitem{LandsmanFlea}
(Klaas)~Landsman, N.P., Reuvers, R.: {A Flea on Schr\"odinger's Cat}.
\newblock Found.\ Phys. \textbf{43}(3), 373--407 (2013).
\newblock \urlprefix\url{http://arxiv.org/abs/1210.2353}

\bibitem{KodeiraThm}
Kodaira, K.: {On K\"ahler varieties of restricted type}.
\newblock Ann.\ Math. \textbf{60}(1), 28--48 (1954)

\bibitem{KontsevichPaper}
Kontsevich, M.: {Deformation Quantization of Poisson Manifolds}.
\newblock Lett.\ Math.\ Phys. \textbf{66}(3), 157--216 (2003).
\newblock \urlprefix\url{http://arxiv.org/abs/q-alg/9709040}

\bibitem{ClassicalInHilbertSpace}
Koopman, B.O.: {Hamiltonian Systems and Transformations in Hilbert Space}.
\newblock {Proc.\ Nat.\ Acad.\ ~Sci.} \textbf{17}(5), 315--318 (1931)

\bibitem{KreiszigBook}
Kreyszig, E.: {Introductory Functional Analysis With Applications}.
\newblock Wiley, U.S.A (1978)

\bibitem{LandsmanBook}
Landsman, N.P.: {Mathematical Topics Between Classical and Quantum Mechanics}.
\newblock Springer Monographs in Mathematics. Springer-Verlag, New York (1998)

\bibitem{PoissonBook}
Laurent-Gengoux, C., Pichereau, A., Vanhaecke, P.: {Poisson Structures}.
\newblock Springer, Heidelberg (2013)

\bibitem{MasanesQM}
Masanes, L., M\"{u}ller, M.P.: {A derivation of quantum theory from physical
  requirements}.
\newblock New J.\ Phys. \textbf{13}, {063,001} (2011).
\newblock \urlprefix\url{http://arxiv.org/abs/1004.1483}

\bibitem{MoyalBracket}
Moyal, J.E.: {Quantum mechanics as a statistical theory}.
\newblock Math.\ Proc.\ Cambridge Phil.\ Soc. \textbf{45}(01), 99--124 (1949)

\bibitem{vonNeuman}
von Neumann, J.: {Zur Operatorenmethode in der klassischen Mechanik}.
\newblock {Proc.\ Nat.\ Acad.\ ~Sci.} \textbf{33}(3), 587--642 (1932)

\bibitem{PironQM}
Piron, C.: {Axiomatique quantique}.
\newblock Helv.\ Phys.\ Acta \textbf{36}, 439--468 (1964)

\bibitem{PR1}
Popescu, S., Rohrlich, D.: {Quantum Nonlocality as an Axiom}.
\newblock {Found.\ Phys.} \textbf{24}(3), 379--385 (1994)

\bibitem{RauPaper}
Rau, J.: {Consistent reasoning about a continuum of hypotheses on the basis of
  finite evidence}.
\newblock arXiv:0706.2274 [quant-ph]  (2001)

\bibitem{SahooPlanck}
Sahoo, D.: {Mixing quantum and classical mechanics and uniqueness of Planck's
  constant}.
\newblock J.\ Phys.\ A:\ Math.\ Gen. \textbf{37}, 997 (2004).
\newblock \urlprefix\url{http://arxiv.org/abs/quant-ph/0301044}

\bibitem{Schlichenmaier}
Schlichenmaier, M.: {Berezin's Coherent States, Symbols and Transform for
  Compact K\"ahler Manifolds}.
\newblock In: P.~Kielanowski, S.T. Ali, A.~Odzijewicz, M.~Schlichenmaier,
  T.~Voronov (eds.) {Geometric Methods in Physics}, Trends in Mathematics, pp.
  101--116. Springer Basel (2013)

\bibitem{SchroeckBook}
Schroeck~Jr., F.E.: {Quantum Mechanics on Phase Space}.
\newblock Kluwer Academic Publishers, Boston (1996)

\bibitem{SegalAxioms}
Segal, I.E.: {Postulates for general quantum mechanics}.
\newblock {Ann.\ Math.} \textbf{48}(4), 930--948 (1947)

\bibitem{StoneTheorem}
Stone, M.H.: {Linear Transformations in Hilbert Space. III. Operational Methods
  and Group Theory}.
\newblock Proc.\ Nat.\ Acad.\ Science \textbf{16}(2), 172--175 (1930)

\bibitem{WeylQuantization}
Upmeier, H.: {Weyl quantization on symmetric spaces I. Hyperbolic matrix
  domains}.
\newblock J.\ Funct.\ Anal. \textbf{92}(2), 297--330 (1991)

\bibitem{Vaidman1}
Vaidman, L.: {Many-Worlds Interpretation of Quantum Mechanics}.
\newblock In: E.N. Zalta (ed.) {The Stanford Encyclopedia of Philosophy},
  spring 2014 edn. (2014)

\bibitem{KazunoriWakatsuki}
Wakatsuki, K.: {Symmetrization of the Berezin Star Product and Multiple Star
  Product Method}.
\newblock J.\ Phys.\ A: Math.\ and Gen. \textbf{34}(37), 7701 (2001).
\newblock \urlprefix\url{http://arxiv.org/abs/hep-th/0006096}

\bibitem{WignerFunctions}
Wigner, E.: {On the Quantum Correction For Thermodynamic Equilibrium}.
\newblock Phys. Rev. \textbf{40}(5), 749--759 (1932)

\bibitem{ZurekEnvariance}
Zurek, W.H.: {Probabilities from entanglement, Born's rule from envariance}.
\newblock Phys. Rev. A \textbf{71}, 052,105 (2005).
\newblock \urlprefix\url{http://arxiv.org/abs/quant-ph/0405161}

\bibitem{ZurekDarwin}
Zurek, W.H.: {Quantum Darwinism}.
\newblock Nat.\ Phys \textbf{5}(3), 181--188 (2009).
\newblock \urlprefix\url{http://arxiv.org/abs/0903.5082}

\end{thebibliography}

\end{document}